\documentclass[12pt]{iopart}

\usepackage{graphicx}

\usepackage{iopams}  
\expandafter\let\csname equation*\endcsname\relax
\expandafter\let\csname endequation*\endcsname\relax
\usepackage{amsmath}

\usepackage{siunitx,cite}
\usepackage{color}
\usepackage[font=small,format=plain,labelfont=bf,up,textfont=normal,up]{caption}
\usepackage[breaklinks=true,colorlinks=true,linkcolor=blue,urlcolor=blue,citecolor=blue]{hyperref}

\usepackage{float}
\usepackage{pgfplots}
\usepackage{tikzscale}
\usepackage{filecontents}
\usepackage{
	pst-plot,
	pst-ode
}

\usetikzlibrary{shapes.callouts}
\tikzset{
	level/.style = {
		ultra thick,
		blue,
	},
	connect/.style = {
		dashed,
		red
	},
	notice/.style = {
		draw,
		rectangle callout,
		callout relative pointer={#1}
	},
	label/.style = {
		text width=2cm
	}
}
\usetikzlibrary{arrows,positioning}
\usetikzlibrary{arrows.meta} 
\begin{document}

\title[Quantum Walks of kicked Bose-Einstein condensates]{Quantum Walks of kicked Bose-Einstein condensates}

\author{C Groiseau$^1$, A Gresch$^1$ and S Wimberger$^{1,2,3}$}
\address{$^1$ITP, Heidelberg University, Philosophenweg 12, 69120 Heidelberg, Germany\\
$^2$Dipartimento di Scienze Matematiche, Fisiche ed Informatiche, Universit\`a di Parma, Parco Area delle Scienze 7/A, 43124 Parma, Italy\\
$^3$INFN, Sezione di Milano Bicocca, Gruppo Collegato di Parma, Parma, Italy}
\ead{sandromarcel.wimberger@unipr.it}
\vspace{10pt}
\begin{indented}
\item[\today]
\end{indented}

\begin{abstract}
We analytically investigate the recently proposed and implemented discrete-time quantum walk based on kicked ultra-cold atoms. We show how the internal level structure of the kicked atoms leads to the emergence of a relative light-shift phase immediately relevant for the experimental realization. Analytical solutions are provided for the momentum distribution for both the case of quantum resonance and the near-resonant quasimomenta.\\
\end{abstract}
\noindent{\it Keywords\/}: Atom Optics Kicked Rotor; Quantum Resonance; Hamiltonian Ratchets; Quantum Walks; Bose-Einstein Condensates

\section{Introduction}
\label{introduction}

Quantum walks \cite{walk} are the quantum-mechanical analogue to classical random walks. The quantum nature of the walkers leads to interference effects in the probability distribution that may have practical applications in the field of quantum information \cite{univprim} or quantum metrology. A variety of different schemes and experimental implementations have been presented, see e.g. \cite{meschede,qwalks}. Here we take a closer look at the recently developed scheme for quantum walks in momentum space \cite{qrw, qrwex}.

\begin{figure}[H]
	\centering
	\includegraphics[width=\textwidth]{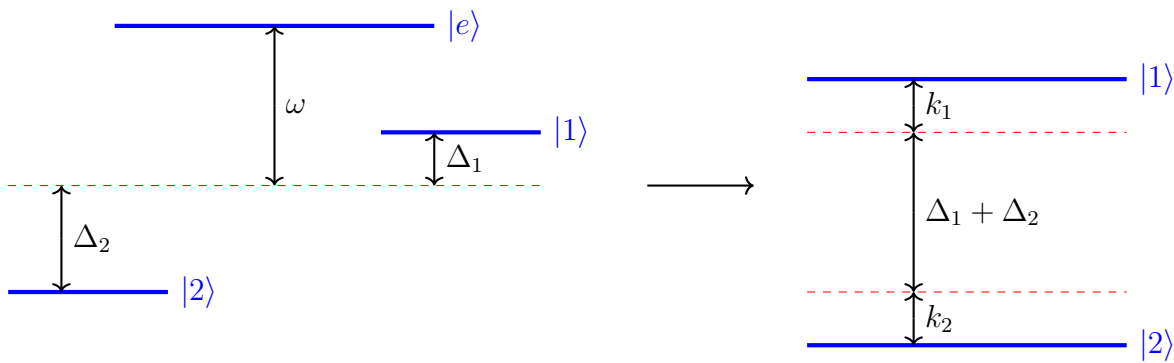}
	\caption{Schematic representation of the system during the kick as an atomic three-level system in $\Lambda$-configuration (left) and of the effective system as an atomic two-level system (right). The adiabatic elimination of the excited state creates an additional light-shift energy difference between the levels, indicated by the two effective kick strengths $k_1$ and $k_2$.}
		\label{fig1}
\end{figure}

The experiment consists in a Bose-Einstein condensate of ultra-cold Rubidium 87 atoms. The two degrees of freedom of this quantum walk scheme are the external centre-of-mass momentum of the atoms and the internal atomic hyperfine states. The two hyperfine levels of the ground state $F=2$ and $F=1$ will be called $|1\rangle$ and $|2\rangle$ in the following. The atoms are kicked by a standing-wave laser of frequency $\omega$ tuned from the excited state manifold $|e\rangle$ between these two ground states (see fig. \ref{fig1} for a schematic representation).

\noindent This setup corresponds to the typical atom optics kicked rotor \cite{hamiltonian} (with the exception of the internal level structure, more on that later) with the rescaled dimensionless Hamiltonian
\begin{eqnarray}
H=\frac{p^2}{2}+k\cos\theta\sum_{j=1}^{T}\delta(t-j\tau),
\end{eqnarray}
where $p$ is the momentum, $\theta$ the periodic position, $k$ the kick strength and $\tau$ the period of the kicks and the kick strength
\begin{eqnarray}
\label{kickstrength}
k=\frac{\Omega^2\tau_p}{8\Delta},
\end{eqnarray}
which can be computed from the Rabi frequency $\Omega$, the finite duration of the kick pulse $\tau_p$ and the detuning of the laser $\Delta$. The Hamiltonian only couples momentum states that differ by a multiple of two photonic recoils so that we may separate the momentum in an integer part $n$ and a conserved non-integer part $\beta$ called quasimomentum
\begin{eqnarray}
p=n+\beta.
\end{eqnarray}
The one-cycle Floquet operator is composed of a kick part
\begin{eqnarray}
\label{kick}
	 K=e^{-ik\cos\theta}
\end{eqnarray}
and a free evolution part
\begin{eqnarray}
	 F=e^{-i\tau\frac{ p^2}{2}}.
\end{eqnarray}
In quantum resonance \cite{qr}, a regime where the kick period $\tau$ is chosen in such a way that the free evolution part is just equal to unity. In this case the momentum distribution of the kicked atoms diffuses symmetrically around its initial integer momentum class and displays ballistic expansion. Such ballistic dynamics were studied before in a similar context for symmetric motion \cite{QR_walk_S} (without the directional control) as well as asymmetric motion \cite{QR_walk_AS} (without the additional coin degree of freedom). Here, we are interested in a directed transport to implement the translational motion conditioned on the internal (coin) degree of freedom. One can break this symmetry by engineering quantum ratchet states. These are states that propagate asymmetrically in momentum space when kicked in a atom optics kicked rotor fashion \cite{ratchet,ratchet2}. A simple superposition of multiple integer momentum classes $n$ is such a state, e.g.
\begin{eqnarray}
\label{ratchetinex}
|\psi\rangle=\frac{1}{\sqrt{2}}\left(|n=0\rangle+e^{i\phi}|n=1\rangle\right),
\end{eqnarray}
which displays an average momentum change per kick
\begin{eqnarray}
\label{meanchange}
	\Delta\langle p\rangle=-\frac{k}{2}\sin\phi.
\end{eqnarray}
The relative phase will be fixed to $\phi=\frac{\pi}{2}$ to maximize the effect of the ratchet. Each step of the quantum walks starts with a pulse of the optical lattice kicking the atoms to induce the momentum change. The direction of the average momentum change can be controlled with the sign of $k$, which for a fixed phase, is solely given by the sign of the detuning as shown in (\ref{meanchange}). By tuning the standing-wave laser between the two ground state levels, so that one is negatively and one positively detuned, we achieve different signs in $k$ as described in (\ref{kickstrength}). Effectively our single state kick operator in (\ref{kick}) changes to
\begin{eqnarray}
\label{kickmatrix}
K=\left(\begin{array}{cc}
e^{-ik\cos\theta} & 0 \\
0 & e^{ik\cos\theta} \\
\end{array}\right).
\end{eqnarray}
The two-momentum-class-state (\ref{ratchetinex}) can be generalized to more complex ratchets with $S$ initial momentum classes
\begin{eqnarray}
|\psi\rangle=\frac{1}{\sqrt{S}}\sum_{s}e^{-is\frac{\pi}{2}}|n=s\rangle
\label{initial_state}
\end{eqnarray}
where the index $s$ takes all values of the momentum classes that make up the ratchet, e.g. $0$ and $1$ for the ratchet in (\ref{ratchetinex}) for $\phi=\frac{\pi}{2}$. The internal degree states are addressed by the two-parameter unitary rotation matrix which in the experiment is done by microwaves. We start by creating an equal superposition of both hyperfine states
\begin{eqnarray}
|\Psi\rangle=\frac{1}{\sqrt{2}}\left(|1\rangle+|2\rangle\right)\otimes|\psi\rangle.
\end{eqnarray}
Then after each of these kicks we mix these internal levels by applying the 50:50 beam splitter coin toss
\begin{eqnarray}
	\label{mix}
	 C=\frac{1}{\sqrt{2}}\left(\begin{array}{cc}
	1 & i \\
	i & 1 \\
	\end{array}\right).
\end{eqnarray}
The total momentum distribution is computed from the sum of the momentum distribution of the two ground states
\begin{eqnarray}
	P(n;T)=P_1(n;T)+P_2(n;T).
	\label{totaldistribution}
\end{eqnarray}
 For more details on the realization of the system and some experimental results we refer to \cite{qrwex}.
 
\section{Effective Dynamics during the Kick}
\label{effective_dynamics}

The effective dynamics during the $\delta$-kick are somewhat different from predicted ones in the previous section. The Hamiltonians in this section only take place during the kick but we refrain from explicitly writing the pulse function envelope for reasons of brevity. The dynamics at this time are given by the interaction picture Hamiltonian in dipole and rotating wave approximation. The two levels are assumed to have the same Rabi frequency $\Omega$.


\begin{eqnarray}
	H_{int}&=\frac{\hbar\Omega}{2}\cos\frac{\theta}{2}\left(|1\rangle\langle e|e^{i\Delta_1 t}+|e\rangle\langle 1|e^{-i\Delta_1 t}\right) \nonumber \\
	&+\frac{\hbar\Omega}{2}\cos\frac{\theta}{2}\left(|2\rangle\langle e|e^{-i\Delta_2 t}+|e\rangle\langle 2|e^{i\Delta_2 t}\right).
\end{eqnarray}
This Hamiltonian is still governed by fast oscillating dynamics and the excited state also still plays a role, but it has an harmonic time dependence 
\begin{eqnarray}
H_{har}=\sum_{n=1}^{2}\left( h_ne^{-i\Delta_nt}+ h_n^\dagger e^{i\Delta_nt}\right)
\end{eqnarray}
with
\begin{eqnarray}
	h_1=\frac{\hbar\Omega}{2}\cos\frac{\theta}{2}|1\rangle\langle e|
\end{eqnarray}
and
\begin{eqnarray}
	h_2=\frac{\hbar\Omega}{2}\cos\frac{\theta}{2}|e\rangle\langle 2|.
\end{eqnarray}
\noindent For Hamiltonians of this form James and Jerke have developed a compact formula in \cite{heff} to derive the time-averaged effective dynamics. To be able to apply this formula we need to fulfil a couple of prerequisites. The atom-field interaction needs to be sufficiently weak and take place over a long period. The first one is easily satisfied by the Rabi frequency being a lot smaller than the atomic transition frequencies. Although the kicked rotor assumes a $\delta$-like kick in reality the kick pulse has a finite width in time of a few hundred nanoseconds \cite{qrwex,hamiltonian,ratchet,ratchet2,ratchet_old,bec1,bec2,bose,adv11}, while the life time of the excited state is about 26 nanoseconds. Finally, fast oscillating terms should be negligible. The application of this procedure yields the effective Hamiltonian
\begin{eqnarray}
	H_{\rm eff}&=\sum_{m,n=1}^{2}\frac{1}{2\hbar}\left(\frac{1}{\Delta_m}+\frac{1}{\Delta_n}\right)\left[ h_m^\dagger, h_n\right]e^{i(\Delta_m-\Delta_n)t}\\
	&=\frac{\hbar\Omega^2}{4\Delta_1}\cos^2\frac{\theta}{2}|1\rangle\langle 1|-\frac{\hbar\Omega^2}{4\Delta_2}\cos^2\frac{\theta}{2}|2\rangle\langle 2|.
\end{eqnarray}
These results are similar to what one gets with the normal kicked rotor without any internal level structure. The difference is that now we have terms for each ground state and that the sign difference in the detuning creates different signs for these two. In the standard kicked rotor the next step is to reformulate the squared cosine using the relation
\begin{eqnarray}
	\cos^2\frac{\theta}{2}=\frac{1}{2}\left(\cos\theta+1\right).
\end{eqnarray} 
The constant part on the right hand side is unproblematic when there is only one ground state because, after adiabatically eliminating the excited state, the system is just a one level system where it creates an energy offset that corresponds to a global phase. Here however, because both ground states create such terms with different signs they add up creating a relative 'light-shift' phase $k_1+k_2$
\begin{eqnarray}
K_{{\rm eff}}=\left(\begin{array}{cc}
e^{-ik_1\cos\theta}e^{-ik_1} & 0 \\
0 & e^{ik_2\cos\theta}e^{ik_2} \\
\end{array}\right).
\end{eqnarray}
The effect of the relative phase has to be counteracted with a phase gate
\begin{eqnarray}
	 M=\left(\begin{array}{cc}
		e^{i\frac{\Phi}{2}} & 0 \\
		0 & e^{-i\frac{\Phi}{2}} \\
	\end{array}\right).
\end{eqnarray}
In addition to the light shift phase we also have to account for the dynamical phase shift that comes from the energy difference of the two hyperfine levels
\begin{eqnarray}
\Phi(\Delta_1,\Delta_2)=k_1+k_2+(\Delta_1+\Delta_2)\tau.
\end{eqnarray}
If the laser is not tuned equidistantly from both levels, one can steer the walks \cite{qrw,qrwex}, but most of the time we choose the detunings to be equal in norm
\begin{eqnarray}
\Phi(\Delta,\Delta)=2k+2\Delta\tau.
\end{eqnarray}
This phase compensation is integrated into our coin from \eqref{mix}
\begin{eqnarray}
 C_{\rm eff}=MC=\frac{1}{\sqrt{2}}\left(\begin{array}{cc}
e^{i\frac{\Phi}{2}} & ie^{-i\frac{\Phi}{2}} \\
ie^{i\frac{\Phi}{2}} & e^{-i\frac{\Phi}{2}} \\
\end{array}\right)
\end{eqnarray}
so that we recover the originally predicted time evolution operator in (\ref{kickmatrix}) because
\begin{eqnarray}
	C_{\rm eff}K_{\rm eff}=CK.
\end{eqnarray}
Hence, any realization of the discrete-time quantum walk following the original proposal \cite{qrw} must use the effective coin in order to compensate the here discussed energy shifts.

\section{The Quantum Resonant Case}
\label{resonant_case}

We proceed by constructing the quantum walk with the time evolution operators that make it up in position space, and then in a final step Fourier transform to momentum space to obtain the momentum distribution.

A single step of the quantum walk consists in kicking the atoms and then mixing the internal states so the total time evolution operator is just the product of (\ref{mix}) and (\ref{kickmatrix}):
\begin{eqnarray}
	 U&= C K\\
	&=\frac{1}{\sqrt{2}}\left(\begin{array}{cc}
	e^{-ik\cos\theta} & ie^{ik\cos\theta} \\
	ie^{-ik\cos\theta} & e^{ik\cos\theta} \\
	\end{array}\right).
\end{eqnarray}
In order to get the final momentum distribution of a walk with a number $T$ of steps we need to find the $T$-th power of the preceding matrix.
\begin{eqnarray}
	 U^T=\left(\frac{1}{\sqrt{2}}\right)^T\left(\begin{array}{cc} A_1^{(T-1)}(k) & A_2^{(T-1)}(k) \\ A_3^{(T-1)}(k) & A_4^{(T-1)}(k) \end{array}\right).
	 \label{ideal_QW_operator}
\end{eqnarray}
The index of the matrix entries $A_i$ are different than the power of the matrix as we are going to link them to recursive polynomials which are usually described with an index indicating their power $N=T-1$ which is shifted by one. The idea here is to express the total time evolution as a polynomial in kick operators $e^{\pm ik\cos\theta}$ in position space which can in the following be translated to momentum space using the relation \cite{mathbook}
\begin{eqnarray}
\label{besselmom}
\int_{0}^{2\pi}e^{in\theta}e^{ik\cos\theta}d\theta=2\pi i^n J_n(k).
\end{eqnarray}
By taking a look at the first few powers of this matrix we notice that the diagonal and off-diagonal matrix elements are identical except for a sign in $k$
\begin{eqnarray}
\label{induction}
A_1^{(N)}(-k)&=A_4^{(N)}(k)\\
\label{induction2}
A_2^{(N)}(-k)&=A_3^{(N)}(k).
\end{eqnarray}
Moreover, we notice is that the matrix entries are constituted by recursive polynomials $p_i$
\begin{eqnarray}
\label{key}
A_1^{(N)}(z)&=e^{-ik\cos\theta}p_1^{(N)}(z)\\
A_2^{(N)}(z)&=ie^{ik\cos\theta}p_2^{(N)}(z)
\end{eqnarray}
in the variable
\begin{eqnarray}
z=e^{-ik\cos\theta}+e^{ik\cos\theta} \,.
\label{eq:z}
\end{eqnarray}
which follow the recursion relation of Dickson polynomials of the second kind
\begin{eqnarray}
\label{recurrence}
p^{(N)}(z)=zp^{(N-1)}(z)-2p^{(N-2)}(z)
\end{eqnarray}
with the initial conditions
\begin{eqnarray}
p_1^{(0)}(z)&=p_2^{(0)}(z)=1\\
p_1^{(1)}(z)&=\tilde z=e^{-ik\cos\theta}-e^{ik\cos\theta}\\
p_2^{(1)}(z)&=z.
\end{eqnarray}
We solve the recursion in appendix A and may express the polynomials as
\begin{eqnarray}
	p_{1/2}^{(N)}(z)=\sum_{l=0}^{N}a_{l,1/2}e^{ik\cos{\theta}(N-2l)}
\end{eqnarray}
where the coefficient are given by
\begin{eqnarray}
\fl	a_{l,1}
	&=\frac{1}{2^{N}}\sum_{j=0}^{\frac{N}{2}}\left(\binom{N}{2j}-\binom{N}{2j+1}\right)\sum_{m=0}^{l}(-8)^m\binom{j}{m}\binom{N-2m}{l-m} \nonumber \\
\fl	&-\frac{1}{2^{N}}2\sum_{j=0}^{\frac{N}{2}}\binom{N}{2j+1}\sum_{m=0}^{l}(-8)^m\binom{j}{m}\binom{N-2m-1}{l-m} \nonumber \\
\fl	&+\frac{1}{2^{N}}2\sum_{j=0}^{\frac{N}{2}}\binom{N}{2j+1}\sum_{m=0}^{l-1}(-8)^m\binom{j}{m}\binom{N-2m-1}{l-m-1}
	\label{coefficient1} \\
\fl	a_{l,2}
	&=\frac{1}{2^{N}}\sum_{j=0}^{\frac{N}{2}}\binom{N+1}{2j+1}\sum_{m=0}^{l}(-8)^m\binom{j}{m}\binom{N-2m}{l-m}.
\label{coefficient2}
\end{eqnarray}
After applying (\ref{besselmom}) we obtain the final momentum distribution, see \ref{mom_distr} for details:
\begin{eqnarray}
\fl	P(n;T)
	&= \frac{1}{2^{T+1}S}\Bigg[\left(\sum_{l=0}^{N}a_{l,1}\left[\sum_{s}(-1)^sJ_{n-s}((N-2l+1)k)\right]\right)^2 \nonumber \\
\fl	&+\left(\sum_{l=0}^{N}a_{l,2}\left[\sum_{s}(-1)^sJ_{n-s}((N-2l+1)k)\right]\right)^2 \nonumber \\
\fl	&+\left(\sum_{l=0}^{N}a_{l,1}\left[\sum_{s}(-1)^sJ_{n-s}(-(N-2l+1)k)\right]\right)^2 \nonumber \\
\fl	&+\left(\sum_{l=0}^{N}a_{l,2}\left[\sum_{s}(-1)^sJ_{n-s}(-(N-2l+1)k)\right]\right)^2\Bigg].
	\label{momentum_distribution_resonance}
\end{eqnarray}

$S$ and $s$ depend on the initial ratchet state as described around equation (\ref{ratchetinex}). Unfortunately, many of the coefficients $a_i$ are in the same order of magnitude and can therefore not be removed. This makes it hard to come up with a good approximation. This formula, however, can easily be expanded for more complex initial quantum ratchet states as long as the relative phase of neighbouring momentum classes stays $\phi=\pm\frac{\pi}{2}$, see the references \cite{ratchet_old,ratchet,ratchet2,qrwex} for the experimental applications of those states. This formula can also be generalized to an arbitrary initial state of the internal degree of freedom $|\Psi\rangle=\left(b_1|1\rangle+b_2|2\rangle\right)\otimes|\psi\rangle$ with $b_i\in\mathbb{R}$ (interesting when doing biased walks \cite{qrwex}) by just multiplying the first and third sum by $b_1^2$ and the second and fourth by $b_2^2$ after removing the global factor $\frac{1}{2}$ that comes from the equal superposition.

\begin{figure}[H]
	\begin{minipage}[b]{\linewidth}
		\centering \includegraphics[width=0.9\textwidth]{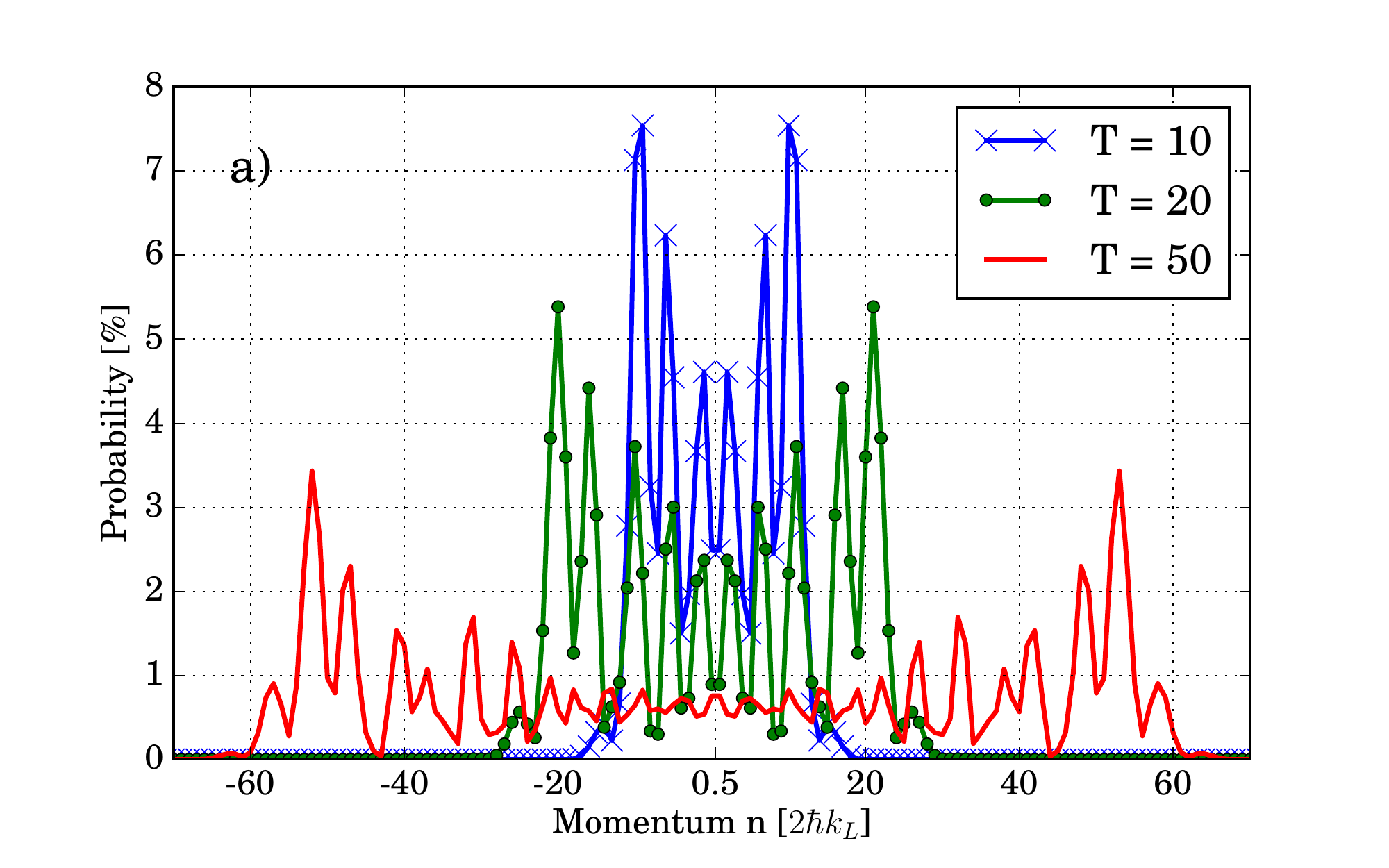}
	\end{minipage} \\
	\begin{minipage}[b]{.5\linewidth}
		\centering \includegraphics[width=0.9\textwidth]{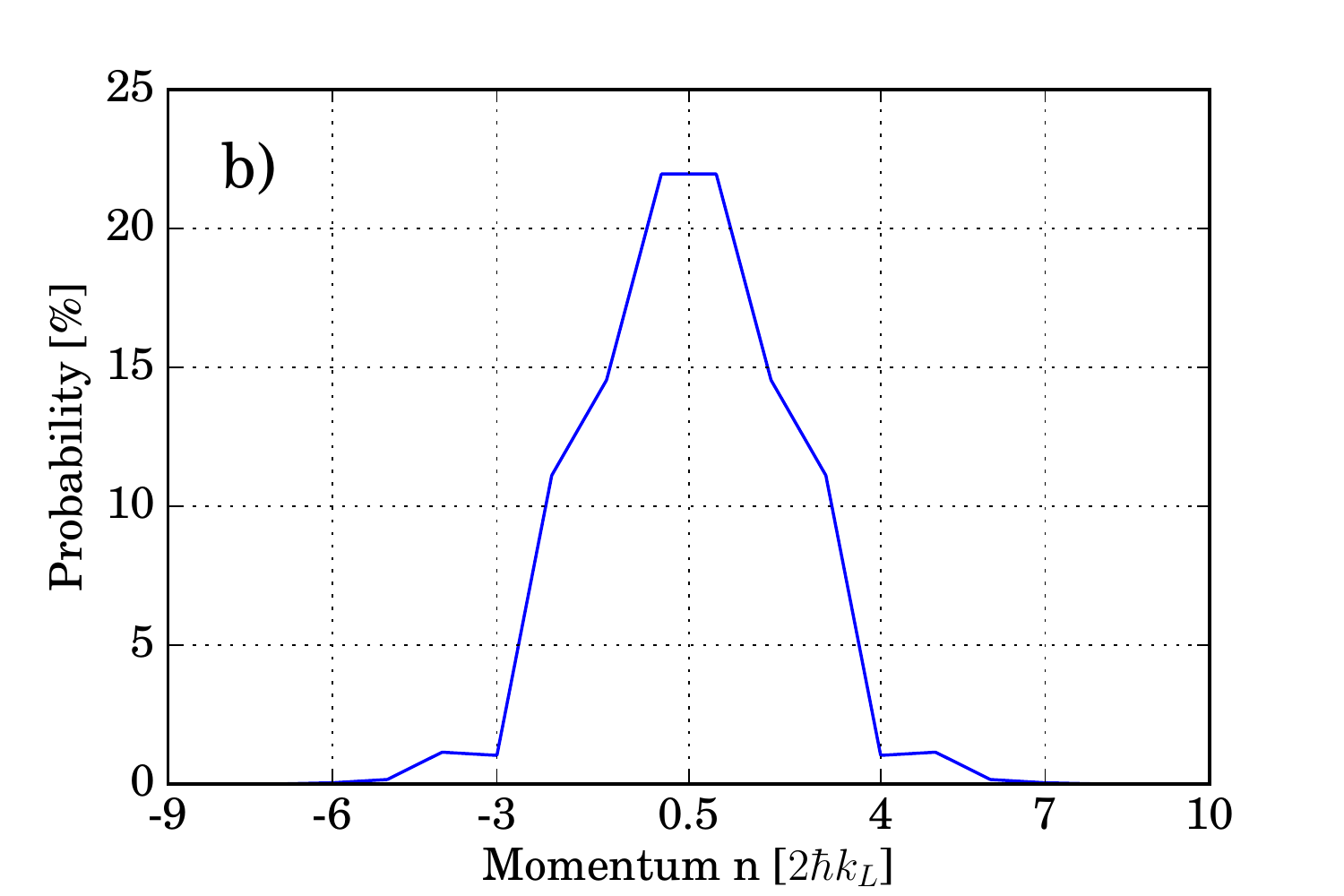}
	\end{minipage}
	\begin{minipage}[b]{.5\linewidth}
		\centering \includegraphics[width=0.9\textwidth]{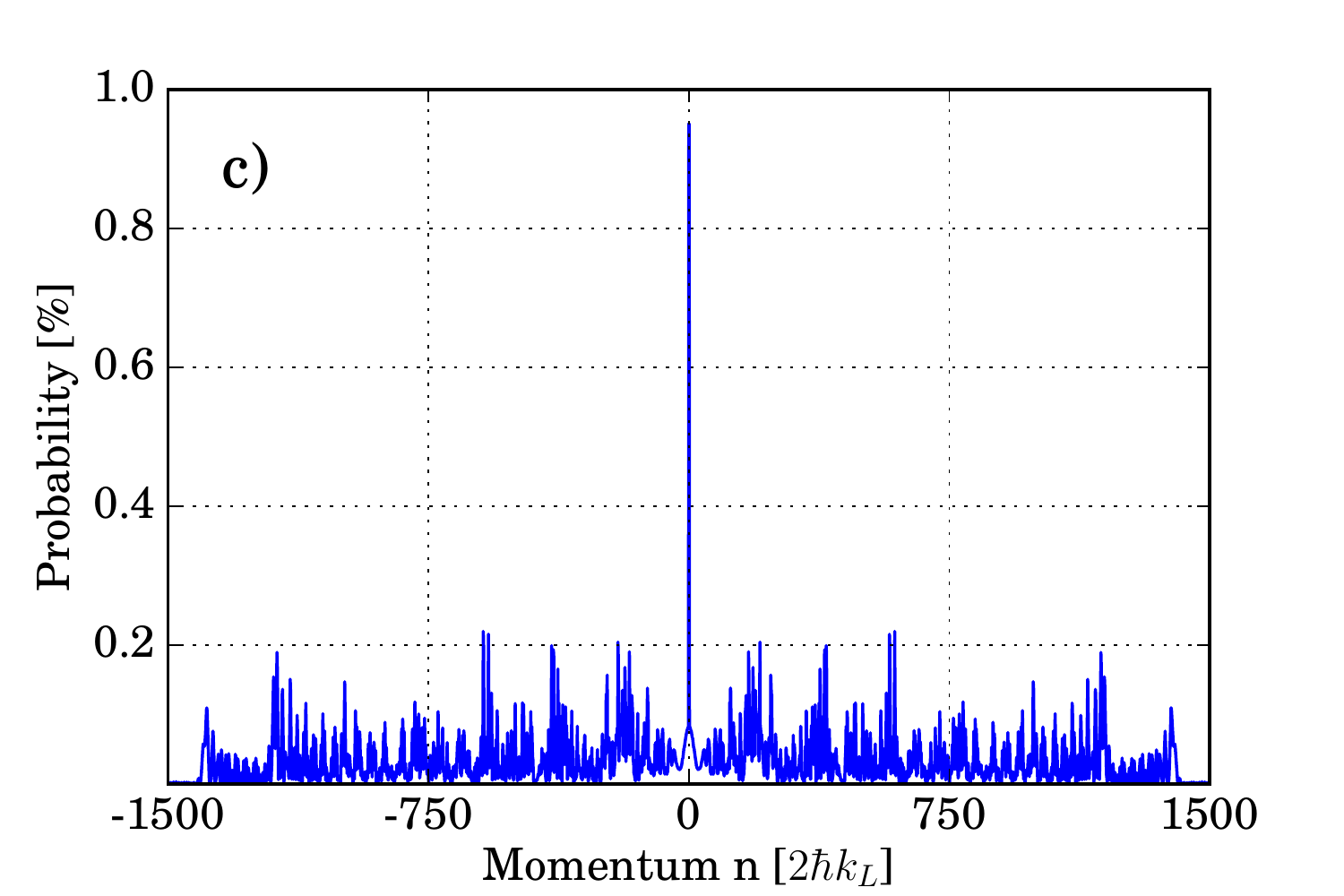}
	\end{minipage}
	\caption{The ideal quantum walk dependence on the kicking strength $k$ and the number of walk steps $T$ taken. a) shows the same walk with $k=2$ at different times and demonstrates the linear growth in the distinctive maxima. If the kicking strength is chosen to be too weak ($k=0.5$ and $T=20$) in b), the walk does not spread at all due to the missing coupling. Conversely, a large coupling ($k=100$ and $T=20$) leads to an effective all-to-all coupling which also destroys the walk.}
	\label{kT-dependence}
\end{figure}

The momentum distribution depends on two main factors: the number of kicks $T$ and the kick strength $k$ visualized in fig. \ref{kT-dependence}. The number of kicks has a small impact on the form of the momentum distribution but the walks show ballistic expansion, the position of the maxima and the standard deviation grow linearly in time. The kick strength has a big impact on the distribution. Small $k$ lead to distributions that never really diffuse over time as the overlap of neighbouring momentum classes in the kick operator is too small. Large $k$ on the other hand lead to 'noisy' distributions as too many momentum classes couple (in a significant manner) to each another. Therefore, $k$ is restricted to a window of $k\in\left[1.0;3.0\right]$ for realistic simulations of discrete-time quantum walks \cite{msc}.

\section{The Near-resonant Case}
\label{near_resonant_case}

The main limitation in the experiments \cite{bec1,bec2} is the finite width in quasimomentum that the Bose-Einstein condensate retains from not being able to be perfectly cooled down to exactly $T=0$K. After a stage of free expansion, the condensate is well approximated by a distribution of Gaussian shape in quasimomenta \cite{bose}. This means that not all the atoms are in quantum resonance conditions \cite{bessel}. These rotors instead of not changing their quantum state during the free evolution evolve with
\begin{eqnarray}
	 F=e^{-i\tau n\beta}
	 \label{free_evolution}
\end{eqnarray}
which shifts the position by $\tau\beta$. This leads to the consecutive step operators depending on $\theta$, $\theta-\tau\beta$, $\theta-2\tau\beta$,... respectively \cite{bessel,adv11}. Since the method we used in quantum resonance relies on $z$ from \eqref{eq:z} always having the same $\theta$-dependency, so that the polynomials always have the same variable, we have to solve the near-resonant case differently here.

The off-resonant rotors have the step-wise time evolution operator
\begin{eqnarray}
	 U_{\beta}&=FCK
	 \label{U_beta} \\
	 &=e^{-i\tau n\beta}\frac{1}{\sqrt{2}}\left(\begin{array}{cc}
	e^{-ik\cos\theta} & ie^{ik\cos\theta} \\
	ie^{-ik\cos\theta} & e^{ik\cos\theta} \\
	\end{array}\right)\\
	&=\frac{1}{\sqrt{2}}\sum_{m\in\mathbb{Z}}^{}\sum_{j\in\mathbb{Z}}^{}e^{-i\tau\beta (j+m)}|j+m\rangle\langle j|J_m(k)i^m\left(\begin{array}{cc}
	(-1)^m & i \\
	i(-1)^m & 1 \\
	\end{array}\right)
	 \label{single_step_operator}
\end{eqnarray}
where in the second step we used the Jacobi-Anger expansion \cite{mathbook} to rewrite the operator in momentum space
\begin{eqnarray}
	e^{\pm ik\cos\theta}=\sum_{m\in\mathbb{Z}}^{}i^mJ_m(\pm k)e^{im\theta}=\sum_{m\in\mathbb{Z}}^{}(\pm i)^mJ_m(k)e^{im\theta},
	\label{Jacobi_Auger}
\end{eqnarray}
and rewrote the translation operator in the momentum basis
\begin{eqnarray}
	e^{im\theta}=\sum_{j\in\mathbb{Z}}^{}|j+m\rangle\langle j|.
	\label{translation_operator}
\end{eqnarray}
(\ref{single_step_operator}) couples each momentum class $j$ with all other neighbour classes $j'$ of difference $m = j' - j$ and weight $J_m(k)$. We also accounted for its respective phase due to the free evolution term of (\ref{free_evolution}). 

From now on, any summation over the indices $m$, $m_l$ or $j$ has a summation range over $\mathbb{Z}$. The concatenation of multiple step-wise operators yields (intermediate steps for subsequent calculations can be found in \ref{appendix_beta_calculation})
\begin{eqnarray}
\fl
U_{\beta}^T&=\left(FCK\right)^T \\
\fl
&=\left(\frac{1}{\sqrt{2}}\right)^T\sum_{m_1,...,m_T}^{}i^{\sum_{l=1}^{T}m_l}\left(\prod_{l=1}^{T}J_{m_l}(k)\right) \sum_{j} e^{-i\tau\beta \left((T)j+\sum_{l=1}^{T} (T+1-l)m_l\right)} \nonumber \\
\fl&\times |j+\sum_{l=1}^{T} m_l\rangle\langle j| R_{T}
\label{QW_operator}
\end{eqnarray}
where $R_T=R_T(\{m_l\})$ is a $(2\times 2)$-matrix of the form
\begin{eqnarray}
	R_T=\prod_{l=1}^{T}\left(\begin{array}{cc}
	(-1)^{m_l} & i \\
	i(-1)^{m_l} & 1 \\
	\end{array}\right).
	\label{recursionmatrix}
\end{eqnarray}
Again, (\ref{QW_operator}) adds up the couplings along all time-steps and weights them with a product of Bessel functions $J_{m_l}$ depending on the respective coupling length $m_l$. We now have the exact time-evolution operator of the walk which now - in contrast to (\ref{kick}) and (\ref{free_evolution}) - only acts in momentum space instead of position space like (\ref{ideal_QW_operator}). 

The next step consists in calculating the actual momentum distribution by computing the bracket product of Bessel functions with the same initial state in (\ref{initial_state})
\begin{eqnarray}
\fl
\langle n,1|U_{\beta}^T|\Psi\rangle&=\left(\frac{1}{\sqrt{2}}\right)^{T+1}\frac{i^n e^{-i\tau\beta n}}{ S} \sum_s (-1)^s e^{-i\tau\beta (T-1)s} \sum_{m_1,\dots ,m_{T-2}} e^{-i\tau\beta \sum_{l=1}^{T-2} (T-l)m_l} \nonumber \\
\fl&\times \left( \prod_{l = 1}^{T-2} J_{m_l}(k) \right) \sum_{m_{T-1}}  J_{m_{T-1}}(k) J_{n-s -\sum_{l=1}^{T-2}m_l-m_{T-1}}(k) \nonumber \\
\fl&\times e^{-i\tau\beta m_{T-1}} \times\left(\ast1\right)_T|_{\sum_{l=1}^{T} m_l= n-s}
\label{mom_calculation}
\end{eqnarray}
where $\left(\ast1\right)_T$ is the sum of the upper two matrix elements of $R_T$ and $s$ being the summation index over all initialized momentum classes in (\ref{ratchetinex}) with norm $S$. We extracted the two last Bessel functions (with indices $m_T$ and $m_{T-1}$) to make use of an addition rule in (\ref{Bessel_addition}) for further simplification. The only meddling term is the phase in the last line of (\ref{mom_calculation}). We approximate that for quasimomenta $\beta$ in the vicinity of zero,
\begin{eqnarray}
e^{i\beta n\pi}J_n(z)\approx	J_n(ze^{i\beta\pi}).
\label{Bessel_approximation}
\end{eqnarray}
This approximation is good for all individual momentum classes $n$ except $n=0,1$ which limitates the applicability of the following calculations. The approximation allows us to absorb the free evolution parts into the argument of the respective Bessel function.
\begin{eqnarray}
\fl
\langle n,1|U_{\beta}^T|\Psi\rangle &\approx\left(\frac{1}{\sqrt{2}}\right)^{T+1}\frac{i^n e^{-i\tau\beta n}}{S} \sum_s (-1)^s e^{-i\tau\beta (T-1)s} \nonumber \\
\fl&\times \sum_{m_1,\dots ,m_{T-2}} \left( \prod_{l = 1}^{T-2} J_{m_l}\left[ k\times e^{-i\tau\beta (T-l)}\right] \right) \nonumber \\
\fl&\times \sum_{m_{T-1}}  J_{m_{T-1}}\left(k\times e^{-i\tau\beta} \right) J_{n-s -\sum_{l=1}^{T-2}m_l-m_{T-1}}(k) \times\left(\ast1\right)_T|_{\sum_{l=1}^{T} m_l = n-s}
\label{eq:approx}
\end{eqnarray}
The two Bessel functions in the last line may be combined using the addition theorem
\begin{eqnarray}
J_n(y+z)=\sum_{r\in\mathbb{Z}} J_r(z)J_{n-r}(y).
\label{Bessel_addition}
\end{eqnarray}
This can be done for all summation indices $M = \{ m_1, \dots, m_T \}$, if we account for the entries of the matrix in (\ref{recursionmatrix}). The diagonal entries simply are the two ratchet currents in each respective walk direction. Terms of the form $(-1)^{m_l}$ can be absorbed as a sign in the respective Bessel functions with the same index. The off-diagonal elements describe a turn in the walk direction where an additional phase $i$ is accumulated. As the two internal levels are mixed after each step, the amount of summands doubles with each step and leave $2^T$ summands of the form
\begin{eqnarray}
i^{\alpha_1} \left( -1 \right)^{\sum_{l \in I } m_l}
\end{eqnarray}
where $I$ is one element of the power set of $M$ and $\alpha_1$ denotes the number of turns during one walk within the internal level. Absorbing the $(-1)^{m_l}$-terms into the arguments of the respective Bessel functions and accounting for all $2^T$ summands we obtain
\begin{eqnarray}
\fl	P_{1/2}(n;T)
l	&\approx\frac{1}{2^{T+1} S} \left| \sum_s (-1)^s e^{-i\tau\beta (T-1)s}  \sum_{c \in \{ 0,1 \}^T} i^{\alpha_{1/2} (c)} \right. \nonumber \\
\fl	&\times \left. J_{n-s}\left[ k\sum_{l=0}^{T-1} (-1)^{c_{l+1}} \left( e^{-i\tau\beta} \right)^l\right] \right|^2
	\label{momentum_distribution_QM}
\end{eqnarray}
for each of the internal levels of the momentum distribution in (\ref{totaldistribution}). $S$ and $s$ once again depend on the initial ratchet state as described in (\ref{ratchetinex}) and $c = (c_1, \dots, c_T)^T$. For general quasimomenta $\beta$, this expression is an approximation limited by the validation range of (\ref{Bessel_approximation}) and comes from the last calculation step in \ref{appendix_beta_calculation}, in (\ref{applied_approximation}). We note that in the quantum resonant case (\ref{momentum_distribution_QM}) is an exact formula since the approximation in (\ref{Bessel_approximation}) is not needed in the calculations. \\
Analogously to above (\ref{momentum_distribution_resonance}), the expression can be generalized to an arbitrary initial state of the internal degree of freedom $|\Psi\rangle=\left(b_1|1\rangle+b_2|2\rangle\right)\otimes|\psi\rangle$ with $b_i\in\mathbb{R}$ by replacing one global factor $\frac{1}{2}$ (from the equal superposition) in $P_{1/2}$ with ${b_{1/2}}^2$ respectively.

The general structure of the momentum distribution is the same as in (\ref{momentum_distribution_resonance}). The main difference arises from the extension to near-resonant quasimomenta $\beta$ which adds phases for the Bessel summands. Another difference here is that we did not account for any recursions as done in (\ref{recurrence}) for the resonant case. It is equivalent to identify all Bessel summands in (\ref{momentum_distribution_QM}) with the same argument, i.e. the same effective kicking strength $k_{\rm eff}$, and adding up their prefactors. This is implicitly included in the coefficients of the resonant momentum distribution ((\ref{coefficient1}) and (\ref{coefficient2})).

\begin{figure}[h]
	\centering
			\includegraphics{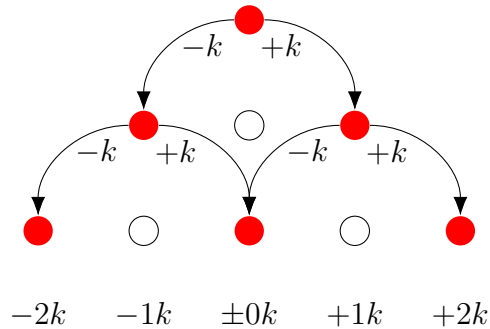}
	\caption{Sketch of the analog to the Galton board, here for $T=2$ kicks in quantum resonance. $2^T = 4$ paths along the board are taken in superposition which correspond to an effective kicking strength of an even multiple of $k$. The two paths with an effective strength of $\pm 0k$ interfere with each other.}
	\label{Galton}
\end{figure}

The summands can be viewed as a walk on the Galton board (quincunx) only now in the domain of the resulting kick strength of the walker (fig. \ref{Galton}). As the particle is $T$-times equally kicked to the left and the right we account for all $2^T$ possible paths on such a board. Going to the left lowers the effective kicking strength by $k$ while going to the right increases it by the same amount. This gives rise to interferences of different paths resulting in the same kicking strength, as for $k_{\rm eff} = 0$ in the sketch. All these paths are considered in (\ref{momentum_distribution_QM}) with their respective phase $i^{\alpha_{1/2}}$ and summed up. Such an addition of effective kicking strengths including the off-resonant quasimomentum was also achieved in \cite{bessel} for the AOKR. 

\begin{center}
\begin{figure}[H]
\begin{minipage}[b]{\linewidth}
\centering \includegraphics[width=0.6\textwidth]{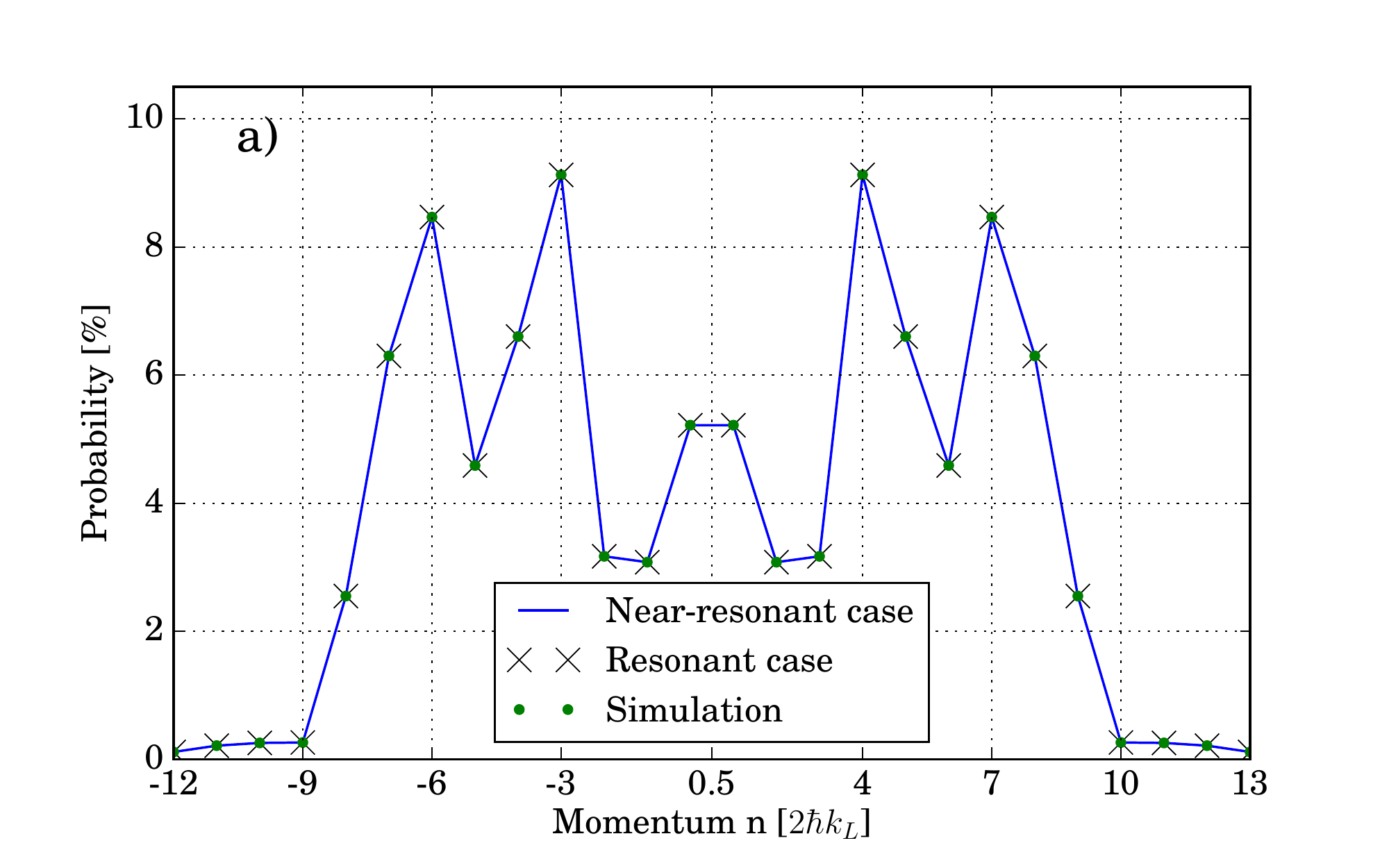}
\end{minipage} \\
\begin{minipage}[b]{\linewidth}
\centering \includegraphics[width=0.6\textwidth]{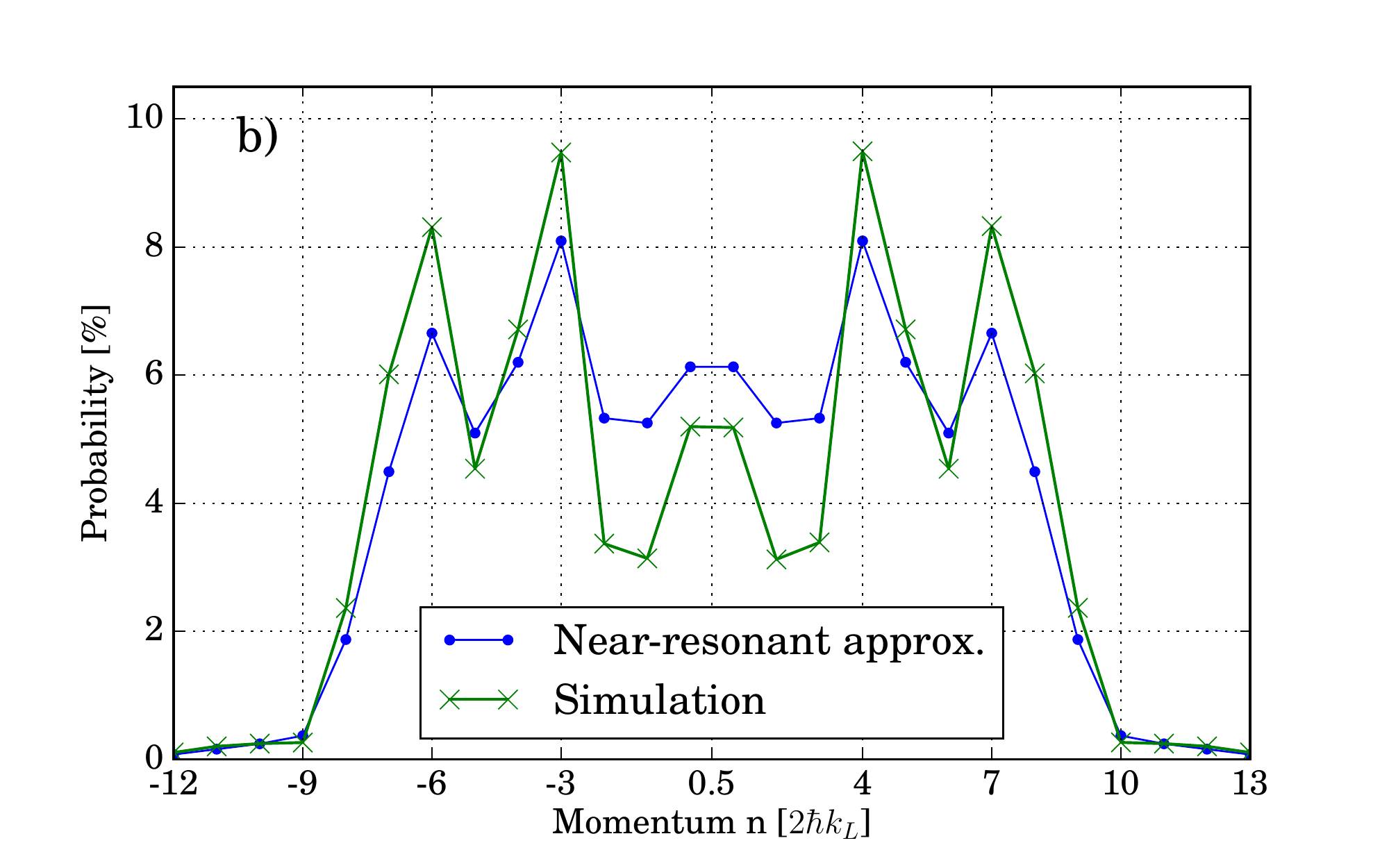}
\end{minipage} \\
\begin{minipage}[b]{\linewidth}
\centering \includegraphics[width=0.6\textwidth]{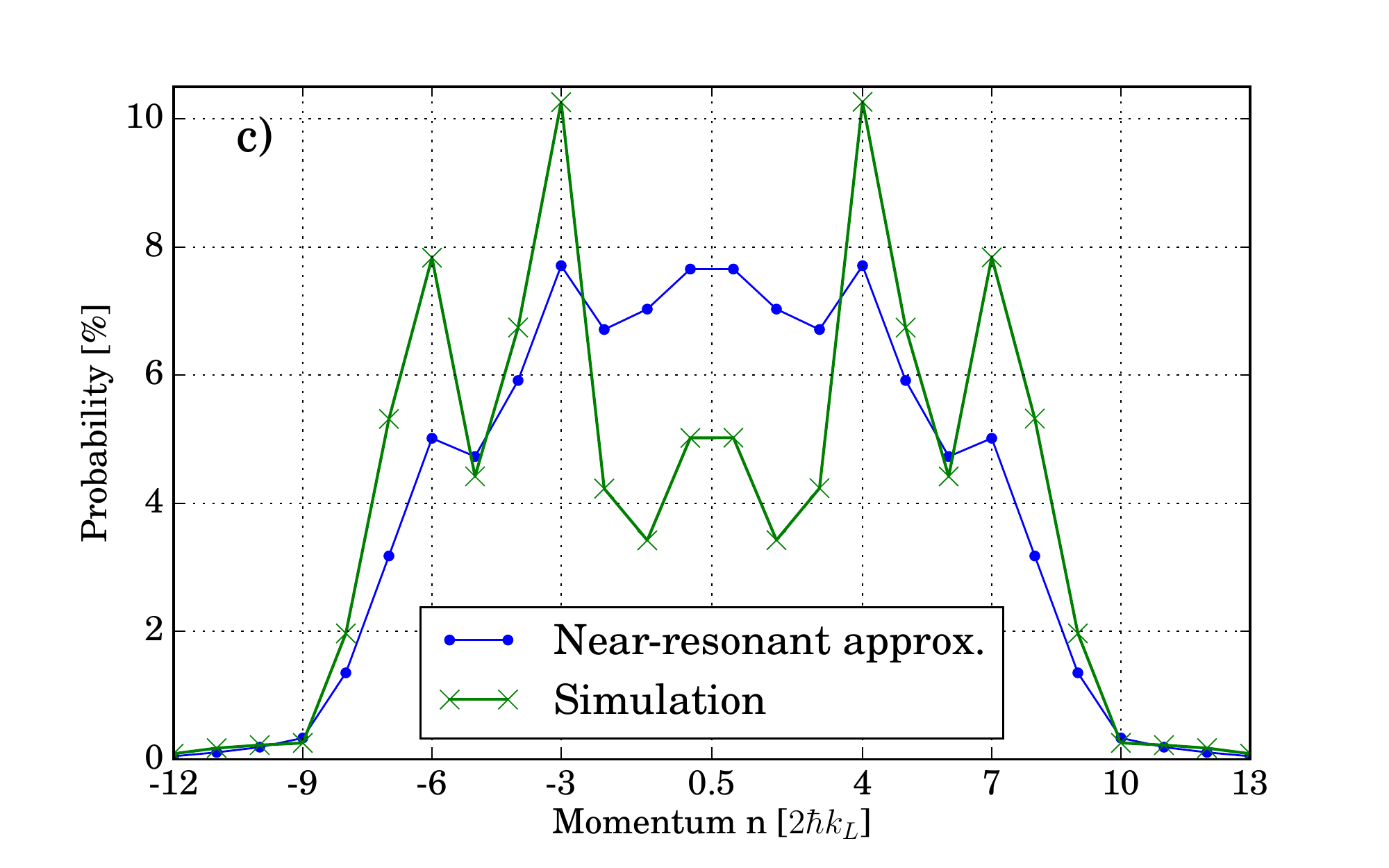}
\end{minipage}
\caption{Comparison of the simulation with the calculated momentum distribution for the resonant case (a) and two near-resonant cases (b,c). The resonant case in (a) corresponds to calculating (\ref{momentum_distribution_resonance}) explicitly, the near-resonant case to (\ref{momentum_distribution_QM}) in the resonance limit $\beta = 0$. Overall good agreement is obtained in (b) for a FWHM of $0.5\%$, where the main difference is around the initial momentum classes due to the approximation of (\ref{Bessel_approximation}) for (\ref{momentum_distribution_QM}). For larger FWHM of $1\%$ in the quasimomenta the approximation is no longer valid (c). Grids drawn to underline the symmetry of the walk.}
\label{QM_simulation_comparison}
\end{figure}
\end{center}

We finally comment on the validity of the approximation with fig. \ref{QM_simulation_comparison}. Since (\ref{Bessel_approximation}) is only relevant for off-resonant quasimomenta, we do not expect any deviations between (\ref{momentum_distribution_resonance}) and (\ref{momentum_distribution_QM}) and a simulation of the quantum walk using a quantum map \cite{msc,bsc} in a). For b) and c) we allowed the rotors to have a near-resonant quasimomentum drawn from a Gaussian distribution with a mean of $0$ (the quantum resonance) and a FWHM $\beta_{FWHM}$ of $0.5\%$ and $1\%$ respectively. The main difference in b) is a higher probability around the initial momentum classes (here we chose $s = 0, 1$ as in (\ref{ratchetinex})). The relevant Bessel summands in (\ref{momentum_distribution_QM}) are the ones with an index of $m = 0,1$ for which the approximation is worst. It manifests in a higher probability of what would be expected. This effect becomes worse for quasimomenta being even farther away from quantum resonance. In c) we chose a higher FWHM 
of $1\%$ which barely shows the features of the accompanying simulation. 

Since the highest effective phase in the argument of the Bessel functions in (\ref{momentum_distribution_QM}) grows with
\begin{eqnarray}
\beta_{\rm eff} \propto \beta\times (T-1),
\end{eqnarray}
there is also a temporal constraint on the validity. We therefore estimate a quantitive validity constraint of $\beta_{FWHM} \times T \leq 10\%$. For state-of-the-art experiments \cite{qrwex}, this implies validity of our approximative formula in \eqref{eq:approx} up to about $T=10$ steps of the walk.

\section{Conclusions}
\label{conclusion}

In summary, we have revisited in detail the proposal for a discrete-time quantum walk in momentum space using a Bose-Einstein condensate with two internal degrees of freedom \cite{qrw}. We discussed the relevance and the quantitative value of the light-shift for the actual three-state system needed to implement the two directed currents contributing to the walk. Fully analytical solutions are obtained for the experimental observable, the final momentum distributions, at quantum resonance conditions. First extensions to finite values of quasimomenta (or, more generally, to values which deviate from the resonant quasimomenta, see \cite{bessel}) are given. 

A challenge is now to reduce the here presented formalism to a more transparent form, possibly based on appropriate approximations. Further extensions of our theory should include any off-resonant values of quasimomenta, which are relevant for the computation of possible thermal clouds present in the experiment reported in ref. \cite{qrwex}. In view of the experimental reality, see  \cite{qrwex}, our analysis will be extended to the case of non-equal kick strength in the two arms of our walk interferometer, i.e. $k_1\neq k_2$, and to include the decohering effect of spontaneous emission on the walk. 


\section*{Acknowledgements}
We thank Gil Summy and Mark Sadgrove for many useful discussions. A.G. gratefully acknowledges support of the PROMOS program by the Heidelberg University and DAAD.

\appendix

\section{Calculations: Resonance}

\subsection{Solving the recursion in (\ref{recurrence})}

We start by solving the recursion, i.e. find a non-recursive analytical form for the polynomials. The uniqueness of the definition in (\ref{recurrence}) is guaranteed by the recursion theorem. To solve this homogeneous linear recurrence relation with constant coefficients we substitute an ansatz $p^{(N)}(z)=x^N(z)$ in the recurrence relation
\begin{eqnarray}
	x^N=zx^{N-1}-2x^{N-2},
\end{eqnarray}
which corresponds to the quadratic equation
\begin{eqnarray}
	x^2=zx-2,
\end{eqnarray}
which has the two solutions
\begin{eqnarray}
	x_{1/2}=\frac{z\pm\sqrt{z^2-8}}{2}.
\end{eqnarray}
Because of the linearity of the recurrence the general solution is
\begin{eqnarray}
	p_{1/2}^{(N)}(z)=c_1x_1^N+c_2x_2^N,
\end{eqnarray}
where $c_1$ and $c_2$ have to be chosen so that the starting conditions are fulfilled, which results in:	
\begin{eqnarray}
	\label{formula}
	p_1^{(N)}(z)&=\frac{1}{2}\left(1+\frac{2\tilde z-z}{\sqrt{z^2-8}}\right)\left(\frac{z+\sqrt{z^2-8}}{2}\right)^N \nonumber \\
	&+\frac{1}{2}\left(1-\frac{2\tilde z-z}{\sqrt{z^2-8}}\right)\left(\frac{z-\sqrt{z^2-8}}{2}\right)^N \\
	\label{formula2}
	p_2^{(N)}(z)
	&=\frac{1}{2}\left(1+\frac{z}{\sqrt{z^2-8}}\right)\left(\frac{z+\sqrt{z^2-8}}{2}\right)^N \nonumber \\
	&+\frac{1}{2}\left(1-\frac{z}{\sqrt{z^2-8}}\right)\left(\frac{z-\sqrt{z^2-8}}{2}\right)^N.
\end{eqnarray}

\subsection{Preliminary calculations}
Before we embark on the actual proof of the recursion we will advance two small calculations that will be useful in the proof.
\begin{eqnarray}
z^2-\tilde z^2&=e^{-2ik\cos\theta}+2+e^{2ik\cos\theta}-(e^{-2ik\cos\theta}-2+e^{2ik\cos\theta}) \\
&=4 \\
\fl
(z\tilde z+z^2-8)(z-\tilde z)&=\Big((e^{-ik\cos\theta}+e^{ik\cos\theta})(e^{-ik\cos\theta}-e^{ik\cos\theta}) \nonumber \\
&+(e^{-ik\cos\theta}+e^{ik\cos\theta})^2-8\Big) \nonumber \\
&\times\Big((e^{-ik\cos\theta}+e^{ik\cos\theta})-(e^{-ik\cos\theta}-e^{ik\cos\theta})\Big) \\
&=\left(e^{-2ik\cos\theta}-e^{2ik\cos\theta}+e^{-2ik\cos\theta}+2+e^{2ik\cos\theta}-8\right) \nonumber \\
&\times 2e^{ik\cos\theta} \\
&=4(e^{-ik\cos\theta}-3e^{ik\cos\theta}) \\
&=4\Big(2(e^{-ik\cos\theta}-e^{ik\cos\theta})-(e^{-ik\cos\theta}+e^{ik\cos\theta})\Big) \\
&=4(2\tilde z-z)
\end{eqnarray}

\subsection{Proof of the recursion in (\ref{key})}

The hypothesis, that the matrix entries follow the polynomial form as said in (\ref{key}) is shown via mathematical induction. The base case is trivially true and now we show the inductive step, that if the $N$-th matrix entries $A_1^{(N)}$ and $A_2^{(N)}$ have the polynomial form so  will $A_1^{(N+1)}$ and $A_2^{(N+1)}$:
\begin{eqnarray}
\fl	A_1^{(N+1)}&=e^{-ik\cos\theta}\left(A_1^{(N)}+iA_2^{(N)}\right)\\
\fl	&=e^{-ik\cos\theta}\left(e^{-ik\cos\theta}p_1^{(N)}-e^{ik\cos\theta}p_2^{(N)}\right)
\end{eqnarray}
\begin{eqnarray}
\fl	A_1^{(N+1)}
	&=e^{-ik\cos\theta}\left(\frac{z+\tilde z}{2}p_1^{(N)}-\frac{z-\tilde z}{2}p_2^{(N)}\right)\\
\fl	&=e^{-ik\cos\theta}\left(zp_1^{(N)}-\frac{z-\tilde z}{2}(p_1^{(N)}+p_2^{(N)})\right)\\
\fl	&=e^{-ik\cos\theta}\Bigg[zp_1^{(N)}-\frac{z-\tilde z}{2}\bigg(\frac{1}{2}(1+\frac{2\tilde z-z}{\sqrt{z^2-8}})(\frac{z+\sqrt{z^2-8}}{2})^N \nonumber \\
\fl	&+\frac{1}{2}(1-\frac{2\tilde z-z}{\sqrt{z^2-8}})(\frac{z-\sqrt{z^2-8}}{2})^N+\frac{1}{2}(1+\frac{z}{\sqrt{z^2-8}})(\frac{z+\sqrt{z^2-8}}{2})^N \nonumber \\
\fl	&+\frac{1}{2}(1-\frac{z}{\sqrt{z^2-8}})(\frac{z-\sqrt{z^2-8}}{2})^N\bigg)\Bigg] \\
\fl	&=e^{-ik\cos\theta}\Bigg[zp_1^{(N)}-\frac{z-\tilde z}{2}\bigg(\frac{1}{2}(2+\frac{2\tilde z}{\sqrt{z^2-8}})(\frac{z+\sqrt{z^2-8}}{2})^N \nonumber \\
\fl	&+\frac{1}{2}(2-\frac{2\tilde z}{\sqrt{z^2-8}})(\frac{z-\sqrt{z^2-8}}{2})^N\bigg)\Bigg] \\
\fl	&=e^{-ik\cos\theta}\Bigg[zp_1^{(N)}-\bigg((\frac{z-\tilde z}{2}+\frac{\tilde z(z-\tilde z)}{2\sqrt{z^2-8}})(\frac{z+\sqrt{z^2-8}}{2})^N \nonumber \\
\fl	&+(\frac{z-\tilde z}{2}-\frac{\tilde z(z-\tilde z)}{2\sqrt{z^2-8}})(\frac{z-\sqrt{z^2-8}}{2})^N\bigg)\Bigg] \\
\fl	&=e^{-ik\cos\theta}\Bigg[zp_1^{(N)}-\bigg((\frac{(z-\tilde z)z}{4}+\frac{z\tilde z(z-\tilde z)}{4\sqrt{z^2-8}}+\frac{\sqrt{z^2-8}(z-\tilde z)}{4} \nonumber \\
\fl	&+\frac{\tilde z(z-\tilde z)}{4})(\frac{z+\sqrt{z^2-8}}{2})^{N-1}+(\frac{(z-\tilde z)z}{4}-\frac{z\tilde z(z-\tilde z)}{4\sqrt{z^2-8}} \nonumber \\
\fl	&-\frac{\sqrt{z^2-8}(z-\tilde z)}{4}+\frac{\tilde z(z-\tilde z)}{4})(\frac{z-\sqrt{z^2-8}}{2})^{N-1}\bigg)\Bigg] \\
\fl	&=e^{-ik\cos\theta}\Bigg[-\bigg((\frac{z^2-\tilde z^2}{4}+\frac{(z\tilde z+z^2-8)(z-\tilde z)}{4\sqrt{z^2-8}})(\frac{z+\sqrt{z^2-8}}{2})^{N-1} \nonumber \\
\fl	&+(\frac{z^2-\tilde z^2}{4}-\frac{(z\tilde z+z^2-8)(z-\tilde z)}{4\sqrt{z^2-8}})(\frac{z-\sqrt{z^2-8}}{2})^{N-1}\bigg)+zp_1^{(N)}\Bigg] \\
\fl	&=e^{-ik\cos\theta}\Bigg[zp_1^{(N)}-\bigg((1+\frac{2\tilde z-z}{\sqrt{z^2-8}})(\frac{z+\sqrt{z^2-8}}{2})^{N-1} \nonumber \\
\fl	&+(1-\frac{2\tilde z-z}{\sqrt{z^2-8}})(\frac{z-\sqrt{z^2-8}}{2})^{N-1}\bigg)\Bigg] \\
\fl	&=e^{-ik\cos\theta}(zp_{1}^{(N)}-2p_{1}^{(N-1)})\\
\fl	&=e^{-ik\cos\theta}p_1^{(N+1)}.
\end{eqnarray}

And idem for the other polynomial:
\begin{eqnarray}
\fl	A_2^{(N+1)}
	&=e^{ik\cos\theta}\left(iA_1^{(N)}+A_2^{(N)}\right)\\
\fl	&=e^{ik\cos\theta}\left(ie^{-ik\cos\theta}p_1^{(N)}+ie^{ik\cos\theta}p_2^{(N)}\right)\\
\fl	&=ie^{ik\cos\theta}\left(\frac{z+\tilde z}{2}p_1^{(N)}+\frac{z-\tilde z}{2}p_2^{(N)}\right)\\
\fl	&=ie^{ik\cos\theta}\left(zp_2^{(N)}-\frac{z+\tilde z}{2}(-p_1^{(N)}+p_2^{(N)})\right)\\
\fl	&=ie^{ik\cos\theta}\Bigg[zp_2^{(N)}-\frac{z+\tilde z}{2}\bigg(-\frac{1}{2}(1+\frac{2\tilde z-z}{\sqrt{z^2-8}})(\frac{z+\sqrt{z^2-8}}{2})^N \nonumber \\
\fl	&-\frac{1}{2}(1-\frac{2\tilde z-z}{\sqrt{z^2-8}})(\frac{z-\sqrt{z^2-8}}{2})^N+\frac{1}{2}(1+\frac{z}{\sqrt{z^2-8}})(\frac{z+\sqrt{z^2-8}}{2})^N \nonumber \\
\fl	&+\frac{1}{2}(1-\frac{z}{\sqrt{z^2-8}})(\frac{z-\sqrt{z^2-8}}{2})^N\bigg)\Bigg] \\
\fl	&=ie^{ik\cos\theta}\Bigg[zp_2^{(N)}-\frac{z+\tilde z}{2}\bigg(\frac{1}{2}(\frac{2(-\tilde z+z)}{\sqrt{z^2-8}})(\frac{z+\sqrt{z^2-8}}{2})^N \nonumber \\
\fl	&+\frac{1}{2}(\frac{2(\tilde z-z)}{\sqrt{z^2-8}})(\frac{z-\sqrt{z^2-8}}{2})^N\bigg)\Bigg] \\
\fl	&=ie^{ik\cos\theta}\Bigg[zp_2^{(N)}-\bigg(\frac{z^2-\tilde z^2}{2\sqrt{z^2-8}}(\frac{z+\sqrt{z^2-8}}{2})^N \nonumber \\
\fl	&+\frac{\tilde z^2-z^2}{2\sqrt{z^2-8}}(\frac{z-\sqrt{z^2-8}}{2})^N\bigg)\Bigg] \\
\fl	&=ie^{ik\cos\theta}\Bigg[zp_2^{(N)}-\bigg((\frac{z(z^2-\tilde z^2)}{4\sqrt{z^2-8}}+\frac{z^2-\tilde z^2}{4})(\frac{z+\sqrt{z^2-8}}{2})^{N-1} \nonumber \\
\fl	&+(\frac{z(-z^2+\tilde z^2)}{4\sqrt{z^2-8}}-\frac{-z^2+\tilde z^2}{4})(\frac{z-\sqrt{z^2-8}}{2})^{N-1}\bigg)\Bigg] \\
\fl	&=ie^{ik\cos\theta}\Bigg[zp_2^{(N)}-\bigg((1+\frac{z}{\sqrt{z^2-8}})(\frac{z+\sqrt{z^2-8}}{2})^{N-1} \nonumber \\
\fl	&+(1-\frac{z}{\sqrt{z^2-8}})(\frac{z-\sqrt{z^2-8}}{2})^{N-1}\bigg)\Bigg] \\
\fl	&=ie^{ik\cos\theta}(zp_{2}^{(N)}-2p_{2}^{(N-1)})\\
\fl	&=ie^{ik\cos\theta}p_2^{(N+1)}.
\end{eqnarray}

\subsection{Rewriting of the polynomials into a more accessible form}

Now that we have shown that the polynomials correctly represent the matrix entries of the time evolution operator we rewrite these polynomials in $z$ into polynomials in kick operators $e^{ik\cos\theta}$:

\begin{eqnarray}
\label{rewritten}
\fl	p_1^{(N)}(z)
	&=\frac{1}{2}(1+\frac{2\tilde z-z}{\sqrt{z^2-8}})(\frac{z+\sqrt{z^2-8}}{2})^N+\frac{1}{2}(1-\frac{2\tilde z-z}{\sqrt{z^2-8}})(\frac{z-\sqrt{z^2-8}}{2})^N\\
\fl	&=\frac{1}{2^{N+1}}\bigg[(1+\frac{2\tilde z-z}{\sqrt{z^2-8}})(z+\sqrt{z^2-8})^N
	+(1-\frac{2\tilde z-z}{\sqrt{z^2-8}})(z-\sqrt{z^2-8})^N\bigg] \\
\fl	&=\frac{1}{2^{N+1}}\bigg[(z+\sqrt{z^2-8})^N+(z-\sqrt{z^2-8})^N \nonumber \\
\fl	&+\frac{2\tilde z-z}{\sqrt{z^2-8}}\left((z+\sqrt{z^2-8})^N-(z-\sqrt{z^2-8})^N\right)\bigg] \\
\fl	&=\frac{1}{2^{N+1}}\bigg[\sum_{j=0}^{N}\binom{N}{j}z^{N-j}(\sqrt{z^2-8})^j+\sum_{j=0}^{N}\binom{N}{j}z^{N-j}(-\sqrt{z^2-8})^j \nonumber \\
\fl	&+\frac{2\tilde z-z}{\sqrt{z^2-8}}\left(\sum_{j=0}^{N}\binom{N}{j}z^{N-j}(\sqrt{z^2-8})^j-\sum_{j=0}^{N}\binom{N}{j}z^{N-j}(-\sqrt{z^2-8})^j\right)\bigg] \\
\fl	&=\frac{1}{2^{N}}\bigg[\sum_{j=0}^{\frac{N}{2}}\binom{N}{2j}z^{N-2j}(z^2-8)^j \nonumber \\
\fl	&+\frac{2\tilde z-z}{\sqrt{z^2-8}}\sum_{j=0}^{\frac{N}{2}}\binom{N}{2j+1}z^{N-2j-1}(\sqrt{z^2-8})^{2j+1}\bigg] \\
\fl	&=\frac{1}{2^{N}}\bigg[\sum_{j=0}^{\frac{N}{2}}\left(\binom{N}{2j}-\binom{N}{2j+1}\right)z^{N-2j}(z^2-8)^j \nonumber \\
\fl	&+2\sum_{j=0}^{\frac{N}{2}}\binom{N}{2j+1}\tilde{z}z^{N-2j-1}(z^2-8)^{j}\bigg] \\
\fl	&=\frac{1}{2^{N}}\bigg[\sum_{j=0}^{\frac{N}{2}}\sum_{m=0}^{j}\sum_{l=0}^{N-2m}\big(\binom{N}{2j}-\binom{N}{2j+1}\big)\binom{j}{m}\binom{N-2m}{l} \nonumber \\
\fl	&\times e^{ik\cos{\theta}(N-2m-2l)}(-8)^m\bigg] \nonumber \\
\fl	&+\frac{1}{2^{N}}2\bigg[\sum_{j=0}^{\frac{N}{2}}\sum_{m=0}^{j}\sum_{l=0}^{N-2m-1}\binom{N}{2j+1}\binom{j}{m}\binom{N-2m-1}{l} \nonumber \\
\fl	&\times e^{ik\cos{\theta}(N-2m-2l-2)}(-8)^m\bigg] \nonumber \\
\fl	&-\frac{1}{2^{N}}2\bigg[\sum_{j=0}^{N}\sum_{m=0}^{j}\sum_{l=0}^{N-2m-1}\binom{N}{2j+1}\binom{j}{m}\binom{N-2m-1}{l} \nonumber \\
\fl	&\times e^{ik\cos{\theta}(N-2m-2l)}(-8)^m\bigg]
\end{eqnarray}
\begin{eqnarray}
\fl	p_1^{(N)}(z)
	=\sum_{l=0}^{N}a_{l,1}e^{ik\cos{\theta}(N-2l)}.
\end{eqnarray}

In the last step $l$ is replaced by the variable $l\rightarrow l+m$.

\begin{eqnarray}
\fl	p_2^{(N)}(z)
	&=\frac{1}{2}(1+\frac{z}{\sqrt{z^2-8}})(\frac{z+\sqrt{z^2-8}}{2})^N+\frac{1}{2}(1-\frac{z}{\sqrt{z^2-8}})(\frac{z-\sqrt{z^2-8}}{2})^N\\
\fl	&=\frac{1}{2^{N+1}}\bigg[(1+\frac{z}{\sqrt{z^2-8}})(z+\sqrt{z^2-8})^N 
+(1-\frac{z}{\sqrt{z^2-8}})(z-\sqrt{z^2-8})^N\bigg] \\
\fl	&=\frac{1}{2^{N+1}}\bigg[(z+\sqrt{z^2-8})^N+(z-\sqrt{z^2-8})^N \nonumber \\
\fl	&+\frac{z}{\sqrt{z^2-8}}\left((z+\sqrt{z^2-8})^N-(z-\sqrt{z^2-8})^N\right)\bigg] \\
\fl	&=\frac{1}{2^{N+1}}\bigg[\sum_{j=0}^{N}\binom{N}{j}z^{N-j}(\sqrt{z^2-8})^j+\sum_{j=0}^{N}\binom{N}{j}z^{N-j}(-\sqrt{z^2-8})^j \nonumber \\
\fl	&+\frac{z}{\sqrt{z^2-8}}\left(\sum_{j=0}^{N}\binom{N}{j}z^{N-j}(\sqrt{z^2-8})^j-\sum_{j=0}^{N}\binom{N}{j}z^{N-j}(-\sqrt{z^2-8})^j\right)\bigg] \\
\fl	&=\frac{1}{2^{N}}\bigg[\sum_{j=0}^{\frac{N}{2}}\binom{N}{2j}z^{N-2j}(z^2-8)^j \nonumber \\
\fl	&+\frac{z}{\sqrt{z^2-8}}\sum_{j=0}^{\frac{N}{2}}\binom{N}{2j+1}z^{N-2j-1}(\sqrt{z^2-8})^{2j+1}\bigg] \\
\fl	&=\frac{1}{2^{N}}\left[\sum_{j=0}^{\frac{N}{2}}\binom{N}{2j}z^{N-2j}(z^2-8)^j+\sum_{j=0}^{\frac{N}{2}}\binom{N}{2j+1}z^{N-2j}(z^2-8)^{j}\right]\\
\fl	&=\frac{1}{2^{N}}\left[\sum_{j=0}^{\frac{N}{2}}\binom{N+1}{2j+1}z^{N-2j}(z^2-8)^j\right]\\
\fl	&=\frac{1}{2^{N}}\left[\sum_{j=0}^{\frac{N}{2}}\sum_{m=0}^{j}\binom{N+1}{2j+1}\binom{j}{m}z^{N-2m}(-8)^m\right]\\
\fl	&=\frac{1}{2^{N}}\left[\sum_{j=0}^{\frac{N}{2}}\sum_{m=0}^{j}\sum_{l=0}^{N-2m}\binom{N+1}{2j+1}\binom{j}{m}\binom{N-2m}{l}e^{ik\cos{\theta}(N-2m-2l)}(-8)^m\right]\\
\fl	&=\sum_{l=0}^{N}a_{l,2}e^{ik\cos{\theta}(N-2l)}
\end{eqnarray}

\subsection{Calculation of the momentum distribution}
\label{mom_distr}
The final momentum distribution of the walk can be computed by using (\ref{besselmom}). 

\begin{eqnarray}
\fl
P(n;T)&=P_1(n;T)+P_2(n;T)\\
\fl
&=\left[\bigg|\frac{1}{\sqrt{2\pi}}\int_{0}^{2\pi}e^{-in\theta}\langle \theta,1|U^T|\Psi\rangle d\theta\bigg|^2+\bigg|\frac{1}{\sqrt{2\pi}}\int_{0}^{2\pi}e^{-in\theta}\langle \theta,2|U^T|\Psi\rangle d\theta\bigg|^2\right]\\
\fl
&=\Bigg[\bigg|\frac{1}{\sqrt{2\pi}}\int_{0}^{2\pi}e^{-in\theta}\left(\frac{1}{\sqrt{2}}\right)^T\langle \theta,1|\begin{pmatrix} A_1^{(T-1)} & A_2^{(T-1)} \\ A_3^{(T-1)} & A_4^{(T-1)} \end{pmatrix}\nonumber\\
\fl
&\times\frac{1}{\sqrt{2}}\left(|1\rangle+|2\rangle\right)\otimes\frac{1}{\sqrt{S}}\sum_{s}e^{-is\frac{\pi}{2}}|n=s\rangle d\theta\bigg|^2\nonumber\\
\fl
&+\bigg|\frac{1}{\sqrt{2\pi}}\int_{0}^{2\pi}e^{-in\theta}\left(\frac{1}{\sqrt{2}}\right)^T\langle \theta,2|\begin{pmatrix} A_1^{(T-1)} & A_2^{(T-1)} \\ A_3^{(T-1)} & A_4^{(T-1)} \end{pmatrix}\nonumber\\
\fl
&\times\frac{1}{\sqrt{2}}\left(|1\rangle+|2\rangle\right)\otimes\frac{1}{\sqrt{S}}\sum_{s}e^{-is\frac{\pi}{2}}|n=s\rangle d\theta\bigg|^2\Bigg]
\\
\fl
&=\frac{1}{2^{T+1}S}\Bigg[\bigg|\frac{1}{\sqrt{2\pi}}\int_{0}^{2\pi}e^{-in\theta}\left(A_1^{(T-1)}+A_2^{(T-1)}\right)\sum_{s}(-i)^s\langle \theta|n=s\rangle d\theta\bigg|^2\nonumber\\
\fl
&+\bigg|\frac{1}{\sqrt{2\pi}}\int_{0}^{2\pi}e^{-in\theta}\left(A_3^{(T-1)}+A_4^{(T-1)}\right)\sum_{s}(-i)^s\langle \theta|n=s\rangle d\theta\bigg|^2\Bigg]\
\\
\fl
&=\frac{1}{2^{T+1}S}\Bigg[\bigg|\frac{1}{\sqrt{2\pi}}\int_{0}^{2\pi}e^{-in\theta}\left(A_1^{(T-1)}+A_2^{(T-1)}\right)\frac{1}{\sqrt{2\pi}}\sum_{s}(-i)^se^{is\theta} d\theta\bigg|^2\nonumber\\
\fl
&+\bigg|\frac{1}{\sqrt{2\pi}}\int_{0}^{2\pi}e^{-in\theta}\left(A_3^{(T-1)}+A_4^{(T-1)}\right)\frac{1}{\sqrt{2\pi}}\sum_{s}(-i)^se^{is\theta} d\theta\bigg|^2\Bigg]
\\
\fl
&=\frac{1}{2^{T+1}S}\Bigg[\bigg|\frac{1}{2\pi}\int_{0}^{2\pi}\sum_{s}e^{-is\frac{\pi}{2}}e^{-i(n-s)\theta}\nonumber\\
\fl
&\times\left(\sum_{l=0}^{N}a_{l,1}e^{ik\cos{\theta}(N-2l-1)}+i\sum_{l=0}^{N}a_{l,2}e^{ik\cos{\theta}(N-2l+1)}\right) d\theta\bigg|^2\Bigg]\nonumber\\
\fl
&+\bigg|\frac{1}{2\pi}\int_{0}^{2\pi}\sum_{s}e^{-is\frac{\pi}{2}}e^{-i(n-s)}\nonumber\\
\fl
&\times\left(\sum_{l=0}^{N}a_{l,1}e^{-ik\cos{\theta}(N-2l-1)}+i\sum_{l=0}^{N}a_{l,2}e^{-ik\cos{\theta}(N-2l+1)}\right) d\theta\bigg|^2\Bigg]
\end{eqnarray}
\begin{eqnarray}
\fl
&=\frac{1}{2^{T+1}S}\Bigg[\bigg|\sum_{l=0}^{N}\sum_{s}a_{l,1}i^{-(n-s)}(-i)^sJ_{-(n-s)}((N-2l-1)k)\nonumber\\
\fl
&+\sum_{l=0}^{N}\sum_{s}a_{l,2}i^{-(n-s)+1}(-i)^sJ_{-(n-s)}((N-2l+1)k)\bigg|^2\nonumber\\
\fl
&+\bigg|\sum_{l=0}^{N}\sum_{s}a_{l,1}i^{-(n-s)}(-i)^sJ_{-(n-s)}(-(N-2l-1)k)\nonumber\\
\fl
&+\sum_{l=0}^{N}\sum_{s}a_{l,2}i^{-(n-s)+1}(-i)^sJ_{-(n-s)}(-(N-2l+1)k)\bigg|^2\Bigg]\\
\fl
\label{trick}
&=\frac{1}{2^{T+1}S}\Bigg[\bigg|\sum_{l=0}^{N}\sum_{s}a_{l,1}(-1)^{n-s}J_{n-s}((N-2l-1)k)\nonumber\\
\fl
&+\sum_{l=0}^{N}\sum_{s}a_{l,2}i(-1)^{n-s}J_{n-s}((N-2l+1)k)\bigg|^2\nonumber\\
\fl
&+\bigg|\sum_{l=0}^{N}\sum_{s}a_{l,1}(-1)^{n-s}J_{n-s}(-(N-2l-1)k)\nonumber\\
\fl
&+\sum_{l=0}^{N}\sum_{s}a_{l,2}i(-1)^{n-s}J_{n-s}(-(N-2l+1)k)\bigg|^2\Bigg]
\\
\fl
&=\frac{1}{2^{T+1}S}\Bigg[\left(\sum_{l=0}^{N}a_{l,1}(-1)^sJ_{n-s}((N-2l-1)k)\right)^2\nonumber\\
&+\left(\sum_{l=0}^{N}\sum_{s}a_{l,2}(-1)^sJ_{n-s}((N-2l+1)k)\right)^2\nonumber\\
\fl
&+\left(\sum_{l=0}^{N}\sum_{s}a_{l,1}(-1)^sJ_{n-s}(-(N-2l-1)k)\right)^2\nonumber\\
\fl
&+\left(\sum_{l=0}^{N}\sum_{s}a_{l,2}(-1)^sJ_{n-s}(-(N-2l+1)k)\right)^2\Bigg]
\end{eqnarray}
In (\ref{trick}) we used the following relation for Bessel function
\begin{eqnarray}
	J_{-n}(k)=(-1)^nJ_n(k)=J_n(-k).
\end{eqnarray}
This latter result gives \eqref{momentum_distribution_resonance} of the main text.

\section{Calculations: Near-resonance}

\subsection{Step-wise calculations of the respective matrix elements}
\label{appendix_beta_calculation}

We start with calculating the single-step time evolution operator from (\ref{U_beta}) to (\ref{single_step_operator}). In the following, the summation indices $j,j',m,m_l$ are always $\in\mathbb{Z}$.
\begin{eqnarray}
U_\beta &= F C K
= e^{-i\tau\beta n}  \frac{1}{\sqrt{2}}
\begin{pmatrix}
1 & i \\
i & 1
\end{pmatrix}
\begin{pmatrix}
e^{-ik\cos\theta} & 0 \\
0 & e^{ik\cos\theta}
\end{pmatrix} \\
&= \frac{1}{\sqrt{2}} e^{-i\tau\beta n}
\begin{pmatrix}
e^{-ik\cos\theta} & i e^{ik\cos\theta} \\
i e^{-ik\cos\theta} & e^{ik\cos\theta}
\end{pmatrix} \\
&= \frac{1}{\sqrt{2}} e^{-i\tau\beta n} \sum_m i^m J_m(k) e^{im\theta }
\begin{pmatrix}
(-1)^m & i \\
i (-1)^m & 1
\end{pmatrix} \\
&= \frac{1}{\sqrt{2}} \sum_m i^m J_m(k) e^{-i\tau\beta n} \sum_j |j+m\rangle\langle j|
\begin{pmatrix}
(-1)^m & i \\
i (-1)^m & 1
\end{pmatrix} \\
&= \frac{1}{\sqrt{2}} \sum_m i^m J_m(k) \sum_j e^{-i\tau\beta (j+m)} |j+m\rangle\langle j|
\begin{pmatrix}
(-1)^m & i \\
i (-1)^m & 1
\end{pmatrix}
\label{appendix_T_1}
\end{eqnarray}
where in the fourth and fifth step we made use of (\ref{Jacobi_Auger}) and (\ref{translation_operator}) respectively. \\
Four two consecutive walk steps we obtain
\begin{eqnarray}
\fl
U^2_\beta &= \Bigg[ \frac{1}{\sqrt{2}} \Bigg.
\begin{pmatrix}
1 & i \\
i & 1
\end{pmatrix}
e^{-i\tau\beta n}
\begin{pmatrix}
e^{-ik\cos\theta} & 0 \\
0 & e^{ik\cos\theta}
\end{pmatrix} \Bigg. \Bigg]^2\\
\fl
&= \frac{e^{-i\tau\beta n}}{\sqrt{2}^2}  \sum_{m_2,m_1} i^{m_1+m_2} J_{m_1}(k) J_{m_2}(k) \sum_{j',j} e^{-i\tau\beta (j+m_1)} \nonumber \\
\fl&\times \langle j'+m_2| \underbrace{\langle j'|j+m_1\rangle}_{= \delta_{j',j+m_1}} \langle j|
\begin{pmatrix}
(-1)^{m_2} & i \\
i (-1)^{m_2} & 1
\end{pmatrix}
\begin{pmatrix}
(-1)^{m_1} & i \\
i (-1)^{m_1} & 1
\end{pmatrix} \\
\fl&= \frac{1}{\sqrt{2}^2}  \sum_{m_2,m_1} i^{m_1+m_2} J_{m_1}(k) J_{m_2}(k) \sum_{j} e^{-i\tau\beta (2j+2m_1+m_2)} |j+m_1+m_2\rangle\langle j| \nonumber \\
\fl&\times
\begin{pmatrix}
(-1)^{m_1}\left[ (-1)^{m_2} -1 \right] & i \left[ (-1)^{m_2} +1 \right] \\
i (-1)^{m_1}\left[ (-1)^{m_2} +1 \right] & -\left[ (-1)^{m_2} -1 \right]
\end{pmatrix}. \label{appendix_T_2}
\end{eqnarray}

\subsection{Proof of (\ref{QW_operator})}

The start of the proof by induction is already given in (\ref{appendix_T_1}). Doing the induction step $T \mapsto T+1$
\begin{eqnarray}
\fl	U^{T+1}_\beta
	&= \Bigg[ \frac{1}{\sqrt{2}}  \sum_{m_{T+1}} i^{m_{T+1}} J_{m_{T+1}}(k) \sum_{j'} e^{-i\tau\beta n} |j'+m_{T+1}\rangle\langle j'| R_T \Bigg] \times U^T_\beta \\
\fl	&= \frac{1}{\sqrt{2}^{T+1}}  \sum_{m_1,\dots ,m_{T+1}}  i^{\sum_{l=1}^T m_l} \left( \prod_{l = 1}^T J_{m_l}(k) \right) i^{m_{T+1}} J_{m_{T+1}}(k) \nonumber \\
\fl	&\times\sum_{j',j} e^{-i\tau\beta n} e^{-i\tau\beta \left(Tj+\sum_{l=1}^T (T+1-l)m_l\right)}
	|j'+m_{T+1}\rangle\underbrace{\langle j'|j+\sum_{l=1}^T m_l\rangle}_{= \delta_{j',j+\sum_{l=1}^T m_l}}\langle j| R_{T+1} \\
\fl	&= \frac{1}{\sqrt{2}^{T+1}}  \sum_{m_1,\dots ,m_{T+1}}  i^{\sum_{l=1}^{T+1} m_l} \left( \prod_{l = 1}^{T+1} J_{m_l}(k) \right) \sum_{j} e^{-i\tau\beta \left(j+\sum_{l=1}^{T+1} m_l\right)} \nonumber \\
\fl	&\times e^{-i\tau\beta \left(Tj+\sum_{l=1}^T (T+1-l)m_l\right)}
	|j+\sum_{l=1}^{T+1} m_l\rangle\langle j| R_{T+1} \\
\fl	&= \frac{1}{\sqrt{2}^{T+1}}  \sum_{m_1,\dots ,m_{T+1}}  i^{\sum_{l=1}^{T+1} m_l} \left( \prod_{l = 1}^{T+1} J_{m_l}(k) \right) 
	\sum_{j} e^{-i\tau\beta \left((T+1)j+\sum_{l=1}^{T+1} (T+2-l)m_l\right)} \nonumber \\
\fl&\times	|j+\sum_{l=1}^{T+1} m_l\rangle\langle j| R_{T+1}
\end{eqnarray}
with the $(2 \times 2)$-matrix $R_T$ as defined in (\ref{recursionmatrix}).

\subsection{Calculation of the momentum distribution}

The calculation for a single walk step, i.e. $T=1$, is rather straight-forward:
\begin{eqnarray}
\fl	\langle n, 1 | U_\beta| \Psi\rangle &= \langle n|
	\begin{pmatrix}
		1 & 0
	\end{pmatrix}
	\frac1{\sqrt{2}} \sum_{m} i^{m}
	J_{m}(k) \sum_j e^{-i\tau\beta (j+m)} |j+m\rangle\langle j| \nonumber \\
\fl	&\times R_1\times\frac1{\sqrt{S}} \sum_s (-i)^s |s\rangle
	\begin{pmatrix}
		1 \\ 1
	\end{pmatrix} \\
\fl	&= \frac1{\sqrt{2S}} \sum_{m} i^{m} J_{m}(k) \times \left[(-1)^m+i\right] \sum_j e^{-i\tau\beta (j+m)} \nonumber \\
\fl	&\times \sum_s (-i)^s \langle n|j+m\rangle\underbrace{\langle j|s\rangle}_{=\delta_{j,s}}
\end{eqnarray}
where the summation index $s$ once more denotes the initialized momentum classes with norm $S$ in (\ref{ratchetinex}).
\begin{eqnarray}
\fl	\langle n, 1 | U_\beta | \Psi\rangle
	&= \frac1{\sqrt{2S}} \sum_{m} i^{m} J_{m}(k) \left[(-1)^m+i\right] \sum_s e^{-i\tau\beta (s+m)} (-i)^s \underbrace{\langle n|s+m\rangle}_{=\delta_{m,n-s} \atop =\delta_{n,s+m}} \\
\fl	&= \frac1{\sqrt{2S}} \sum_{s} i^{n-s} (-i)^s e^{-i\tau\beta n} J_{n-s}(k) \left[(-1)^{n-s}+i\right] \\
\fl	&= \frac{i^n e^{-i\tau\beta n}}{\sqrt{2S}} \sum_{s} (-1)^s \left[ (-1)^{n-s} J_{n-s}(k)+i J_{n-s}(k)\right] \\
\fl	&= \frac{i^n e^{-i\tau\beta n}}{\sqrt{2S}} \sum_{s} (-1)^s \left[ J_{n-s}(-k)+i J_{n-s}(k)\right] \\
\fl	\langle n, 2 | U_\beta | \Psi\rangle &= \frac1{\sqrt{2S}} \sum_{s} i^{n-s} (-i)^s e^{-i\tau\beta n} J_{n-s}(k) \left[i(-1)^{n-s}+1\right] \\
\fl	&= \frac{i^n e^{-i\tau\beta n}}{\sqrt{2S}} \sum_{s} (-1)^s \left[ iJ_{n-s}(-k)+ J_{n-s}(k)\right]
\end{eqnarray}
yielding
\begin{eqnarray}
\fl	P_1 (n;T) &= \left| \langle n, 2 | U_\beta | \Psi\rangle\right|^2 = \frac1{2S} \left|\sum_{s} (-1)^s \left[ iJ_{n-s}(-k)+ J_{n-s}(k)\right] \right|^2 \\
\fl	P_2 (n;T) &= \left| \langle n, 1 | U_\beta | \Psi\rangle\right|^2 = \frac1{2S} \left|\sum_{s} (-1)^s \left[ J_{n-s}(-k)+i J_{n-s}(k)\right] \right|^2.
\label{appendix_matrix_element_T=1}
\end{eqnarray}

A similar result as above is desired to obtain for a higher kick count and - with the described approximation in (\ref{Bessel_approximation}) - is achieved in the following. We demonstrate the calculations for $T=2$ using the approximation and the addition rule for the Bessel functions, (\ref{Bessel_addition}).
\begin{eqnarray}
\fl	\langle n, 1 | U^2_\beta | \Psi\rangle &= \langle n|
	\begin{pmatrix}
		1 & 0
	\end{pmatrix}
	\frac12 \sum_{m_1,m_2} i^{m_1+m_2}
	J_{m_1}(k) J_{m_2}(k) e^{-i\tau\beta (2j+2m_1+m_2)} \nonumber \\
\fl	&\times\sum_j |j+m_1+m_2\rangle\langle j| R_2\times\frac1{\sqrt{S}} \sum_s (-i)^s |s\rangle
	\begin{pmatrix}
		1 \\ 1
	\end{pmatrix} \\
\fl	&= \frac1{2\sqrt{S}} \sum_{m_1,m_2} i^{m_1+m_2} \left(\ast 1\right)_2 J_{m_1}(k) J_{m_2}(k) e^{-i\tau\beta (2j+2m_1+m_2)} \nonumber \\
\fl	&\times\sum_j \langle n | j+m_1+m_2\rangle\times \sum_s (-i)^s \langle j | s\rangle
\end{eqnarray}
where $\left(\ast 1\right)_2$ denotes the sum of the two upper matrix elements of $R_2$ in (\ref{appendix_T_2}) and is abbreviated for short-hand notations.
\begin{eqnarray}
\fl	\langle n, 1 | U^2_\beta | \Psi\rangle &= \frac1{2\sqrt{S}} \sum_{m_1,m_2} i^{m_1+m_2} \left(\ast 1\right)_2 J_{m_1}(k) J_{m_2}(k) \nonumber \\
\fl	&\times\sum_s (-i)^s e^{-i\tau\beta (2s+2m_1+m_2)} \underbrace{\langle n|s+m_1+m_2\rangle}_{=\delta_{n-s,m_1+m_2} \atop =\delta_{m_2,n-s-m_1}} \\
\fl	&= \frac1{2\sqrt{S}} \sum_s i^{n-s} (-i)^s \sum_{m_1} \left(\ast 1\right)_2 |_{m_2 \atop = n-s-m_1} \nonumber \\
\fl	&\times J_{m_1}(k) J_{n-s-m_1}(k) e^{-i\tau\beta (2s+m_1+n-s)} \\
\fl	&= \frac{i^n e^{-i\tau\beta n}}{2\sqrt{S}} \sum_s (-1)^s e^{-i\tau\beta s} \sum_{m_1} \left(\ast 1\right)_2 |_{m_2 \atop = n-s-m_1} \nonumber \\
\fl	&\times J_{m_1}(k) J_{n-s-m_1}(k) e^{-i\tau\beta m_1}
\end{eqnarray}
The approximation of (\ref{Bessel_approximation}) results in
\begin{eqnarray}
\fl	\langle n, 1 | U^2_\beta | \Psi\rangle &\approx \frac{i^n e^{-i\tau\beta n}}{2\sqrt{S}} \sum_s (-1)^s e^{-i\tau\beta s} \sum_{m_1} \left(\ast 1\right)_2 |_{m_2 \atop = n-s-m_1} \nonumber \\
\fl	&\times J_{m_1}(k\times e^{-i\tau\beta}) J_{n-s-m_1}(k) \\
\fl	&=\frac{i^n e^{-i\tau\beta n}}{2\sqrt{S}} \sum_s (-1)^s e^{-i\tau\beta s} \sum_{m} J_{m}(k\times e^{-i\tau\beta}) J_{n-s-m}(k) \nonumber \\
\fl	&\times \left( \left[ (-1)^{n-s} - (-1)^{m} \right] + i \left[ (-1)^{n-s-m} + 1 \right] \right)
\label{recurrence_input_T=2} \\
\fl	&=\frac{i^n e^{-i\tau\beta n}}{2\sqrt{S}} \sum_s (-1)^s e^{-i\tau\beta s} \Bigg[ (-1)^{n-s} \sum_{m} J_{m}(k\times e^{-i\tau\beta}) J_{n-s-m}(k) \Bigg. \nonumber \\
\fl	&- \sum_{m} J_{m}(-k\times e^{-i\tau\beta}) J_{n-s-m}(k) +i \sum_{m} J_{m}(k\times e^{-i\tau\beta}) J_{n-s-m}(-k) \nonumber \\
\fl	&+ i \sum_{m} J_{m}(k\times e^{-i\tau\beta}) J_{n-s-m}(k) \Bigg. \Bigg] \\
\fl	&=\frac{i^n e^{-i\tau\beta n}}{2\sqrt{S}} \sum_s (-1)^s e^{-i\tau\beta s} \left[ (-1)^{n-s} J_{n-s} \left[ k \left( 1+ e^{-i\tau\beta} \right) \right] \right. \nonumber \\
\fl	&- J_{n-s} \left[ k \left( 1- e^{-i\tau\beta} \right)\right] +i J_{n-s}\left[ k \left( e^{-i\tau\beta}-1 \right) \right] \nonumber \\
\fl	&\left.+ i J_{n-s} \left[ k \left(1+ e^{-i\tau\beta}\right)\right] \right] \\
\fl	&=\frac{i^n e^{-i\tau\beta n}}{2\sqrt{S}} \sum_s (-1)^s e^{-i\tau\beta s} \left[ J_{n-s} \left[ -k \left( 1+ e^{-i\tau\beta} \right) \right] \right. \nonumber \\
\fl	&- J_{n-s} \left[ k \left( 1- e^{-i\tau\beta} \right)\right] +i J_{n-s}\left[ k \left( e^{-i\tau\beta}-1 \right) \right] \nonumber \\
\fl	&\left.+ i J_{n-s} \left[ k \left(1+ e^{-i\tau\beta}\right)\right] \right]
\label{eq:appendix_matrix_element_T=2}
\end{eqnarray}
yielding (analogously for the other internal level)
\begin{eqnarray}
\fl	\left|\langle n, 1 | U^2_\beta | \Psi\rangle\right|^2 &\approx \frac1{4S} \Bigg| \sum_s (-1)^s e^{-i\tau\beta s} \left( J_{n-s} \left[ -k \left( 1+ e^{-i\tau\beta} \right) \right] \right. \Bigg. \nonumber \\
\fl	& - J_{n-s} \left[ k \left( 1- e^{-i\tau\beta} \right)\right] +i J_{n-s}\left[ k \left( e^{-i\tau\beta}-1 \right) \right] \nonumber \\
\fl	&\Bigg.\left. + i J_{n-s} \left[ k \left(1+ e^{-i\tau\beta}\right)\right] \right)\Bigg|^2 \\
\fl	\left|\langle n, 2 | U^2_\beta | \Psi\rangle\right|^2 &\approx \frac1{4S} \Bigg| \sum_s (-1)^s e^{-i\tau\beta s} \left( iJ_{n-s} \left[ -k \left( 1+ e^{-i\tau\beta} \right) \right] \right. \Bigg. \nonumber \\
\fl	& +i J_{n-s} \left[ k \left( 1- e^{-i\tau\beta} \right)\right] - J_{n-s}\left[ k \left( e^{-i\tau\beta}-1 \right) \right] \nonumber \\
\fl	&\Bigg.\left. + J_{n-s} \left[ k \left(1+ e^{-i\tau\beta}\right)\right] \right)\Bigg|^2.
\end{eqnarray}
We again note that for the limiting case of quantum resonance, i.e. $\beta = 0$, the momentum distribution becomes exact again. \\

The calculation for higher walk steps $T > 2$ can be started using (\ref{QW_operator}):
\begin{eqnarray}
\fl	\langle n, 1 | U^T_\beta | \Psi\rangle &= \frac{1}{\sqrt{2}^T \sqrt{S}}  \sum_{m_1,\dots ,m_T} i^{\sum_{l=1}^T m_l} \left( \prod_{l = 1}^T J_{m_l}(k) \right) \nonumber \\
\fl	&\times \sum_{j,s} (-i)^s e^{-i\tau\beta \left(Tj+\sum_{l=1}^T (T+1-l)m_l\right)} \nonumber \\
\fl	&\times\underbrace{\langle n|j+\sum_{l=1}^T m_l\rangle\langle j|s\rangle}_{=\delta_{n-s,\sum m_l}}
	\begin{pmatrix}
		1 & 0
	\end{pmatrix}
	R_T
	\begin{pmatrix}
		1 \\ 1
	\end{pmatrix} \\
\fl	&= \frac{1}{\sqrt{2}^T \sqrt{S}} \sum_s \sum_{m_1,\dots ,m_T \atop \sum m_l = n-s} i^{n-s} (-i)^s \left( \prod_{l = 1}^T J_{m_l}(k) \right) \nonumber \\
\fl	&\times e^{-i\tau\beta \left(Ts+\sum_{l=1}^T (T+1-l)m_l\right)} \times\left(\ast 1\right)_T |_{\sum_{l=1}^{T} m_l = n-s} \\
\fl	&= \frac{i^n}{\sqrt{2}^T \sqrt{S}} \sum_s (-1)^s e^{-i\tau\beta Ts} \sum_{m_1,\dots ,m_{T-1}} e^{-i\tau\beta \sum_{l=1}^{T-1} (T-l)m_l} \nonumber \\
\fl	&\times \left( \prod_{l = 1}^{T-2} J_{m_l}(k) \right) \sum_{m_T} \delta_{\sum_{l=1}^{T} m_l,n-s} J_{m_{T-1}}(k)  \nonumber \\
\fl	&\times\underbrace{J_{m_T}(k)}_{m_T = n-s \atop -m_{T-1}-\sum_{l=1}^{T-2} m_l}\underbrace{e^{-i\tau\beta \sum m_l}}_{= e^{-i\tau\beta (n-s)}} \times\left(\ast 1\right)_T |_{\sum_{l=1}^{T} m_l = n-s}
\end{eqnarray}
\begin{eqnarray}
\fl	\langle n, 1 | U^T_\beta | \Psi\rangle &= \frac{i^n e^{-i\tau\beta n}}{\sqrt{2}^T \sqrt{S}} \sum_s (-1)^s e^{-i\tau\beta (T-1)s} \sum_{m_1,\dots ,m_{T-2}} e^{-i\tau\beta \sum_{l=1}^{T-2} (T-l)m_l} \nonumber \\
\fl	&\times \left( \prod_{l = 1}^{T-2} J_{m_l}(k) \right) \sum_{m_{T-1}}  J_{m_{T-1}}(k) J_{n-s -\sum_{l=1}^{T-2}m_l-m_{T-1}}(k) \nonumber \\
\fl	&\times e^{-i\tau\beta m_{T-1}} \times\left(\ast 1\right)_T |_{\sum_{l=1}^{T} m_l = n-s} \\
\fl	&\approx \frac{i^n e^{-i\tau\beta n}}{\sqrt{2}^T \sqrt{S}} \sum_s (-1)^s e^{-i\tau\beta (T-1)s} \sum_{m_1,\dots ,m_{T-2}} \nonumber \\
\fl	&\times \left( \prod_{l = 1}^{T-2} J_{m_l}\left[ k\times e^{-i\tau\beta (T-l)}\right] \right) \sum_{m_{T-1}}  J_{m_{T-1}}\left(k\times e^{-i\tau\beta} \right) \nonumber \\
\fl	&\times J_{n-s -\sum_{l=1}^{T-2}m_l-m_{T-1}}(k) 
	\times\left(\ast 1\right)_T |_{\sum_{l=1}^{T} m_l = n-s}.
\label{applied_approximation}
\end{eqnarray}
In the last step, the approximation of (\ref{Bessel_approximation}) was applied as before to allow the application of the addition rule in (\ref{Bessel_addition}).

\bibliographystyle{unsrt}

\end{document}